\documentclass[journal,onecolum]{IEEEtran}
\usepackage{amsmath}
\usepackage{graphicx}
\usepackage{textcomp}
\usepackage{xcolor}

\usepackage{subcaption}
\usepackage{epsfig}
\usepackage{epstopdf} 
\usepackage{tikz}
\usepackage[hidelinks]{hyperref}

\usepackage{xcolor}
\usepackage{amssymb}
\ifCLASSINFOpdf
\else

\fi

\begin{document}
%

\title{Biocomputing Model Using Tripartite Synapses Provides Reliable Neuronal Logic Gating with Spike Pattern Diversity}

\author{Giulio Basso and 
        Michael Taynnan Barros,~\IEEEmembership{Member,~IEEE,}
\thanks{G. Basso is with the Department of Mechanical and Aerospace Engineering, Politecnico di Torino, Italy. e-mail: s278524@studenti.polito.it}
\thanks{M. Barros is with the School of Computer Science and Electronic Engineering, University of Essex, Colchester, United Kingdom, and CBIG/Biomeditech, Faculty of Medicine and Health Technology, Tampere University, Finland. e-mail: m.barros@essex.ac.uk.}
}

\maketitle

\begin{abstract}
Biocomputing technologies exploit biological communication mechanisms involving cell-cell signal propagation to perform computations. Researchers recently worked toward realising logic gates made by neurons to develop novel devices such as organic neuroprostheses or brain implants made by cells, herein termed living implants. Several challenges arise from this approach, mainly associated with the stochastic nature and noise of neuronal communication. Since astrocytes play a crucial role in the regulation of neurons activity, there is a possibility whereby astrocytes can be engineered to control synapses favouring reliable biocomputing. This work proposes a mathematical model of neuronal logic gates involving neurons and astrocytes, realising OR and AND gating. We use a shallow coupling of both the Izhikevich and Postnov models to characterise gating responses with spike pattern variability and astrocyte synaptic regulation. Logic operation error ratio and accuracy assess the AND and OR gates' performances at different synaptic Gaussian noise levels. Our results demonstrate that the astrocyte regulating activity can effectively be used as a denoising mechanism, paving the way for highly reliable biocomputing implementations.
\end{abstract}


\IEEEpeerreviewmaketitle

\section{Introduction}
Biological systems, like cells, communicate with each other encoding information by modulating the concentration of specific molecules. Recently communication engineering researchers have considered the possibility of exploiting these cells' communication mechanisms to realise communication systems, termed Molecular communication (MC). Such technology allows the controllable exchange of biological information in order to realise human-designed tasks, including biocomputing \cite{akyildiz2019moving}. Most biocomputing technologies under investigation use DNA as information molecule as well as calcium signalling \cite{tavella2019dna,barros2021engineering}. For instance, particular genes and ions can be transmitted from a bio-transmitter to a bio-receiver, controlling the expression of specific proteins, with the objective of introducing new functionalities or restoring physiologic mechanisms altered by diseases \cite{akyildiz2019moving,kuscu2019transmitter}. This innovative approach also allows the employment of communication engineering and information theory in the biological sciences, biotechnology, and medicine.  MC-based technologies have been investigated as part of promising tools regarding tumour treatments when traditional anti-tumour therapies could fail \cite{veletic2020modeling}. They can be employed for active targeted drug delivery: since tumour cells often can present receptors overexpressed with respect to healthy cells, specific ligand-receptor reactions can be exploited to transport anti-tumour drugs in the tumour site, minimising the toxic effects in the other tissues. 
Several benefits can then result from the use of biocomputing, where these bio-transmitters and bio-receivers can have the power to process the biological information molecules as well in a controlled logical approach. Devices with high biocompatibility can be developed as well \cite{egan2018strategies}, which can directly communicate with the natural biological environment in order to avoid the aggressive response of the immune system with respect to the artificial implant \cite{akyildiz2019moving,barros2021engineering,adonias2020reconfigurable}. Furthermore, communication in biological systems is characterised by an extraordinarily high-efficiency \cite{galluccio2011characterization}. Since molecules are available at the biological nano-scale, time constants, such as diffusion times, are reduced, reagents take less time to diffuse and mix in the reaction space, and so chemical reactions are facilitated \cite{egan2021stochastic,soldner2020survey}.

Biocomputing solutions using mammalian cells, specifically, face many challenges due to intrinsic stochasticity that stops further progression to in-vivo applications. The biological environment is strongly noisy and biological signals coded into molecules are unpredictable \cite{moore2007interfacing}. Since denoising mechanisms are not currently available, the signal-to-noise ratio needs to be increased by sending a large number of information molecules \cite{nakano2012molecular,moore2007interfacing}. Moreover, the natural molecules present in the environment could interfere with the information molecule propagation \cite{egan2018strategies}, and so the resulting communication systems could be affected by a delay \cite{moore2007interfacing}. Additionally, when molecules remain in the environment for a prolonged time they can suffer from degradation, leading to the possible loss of functionality. This can then impair the accuracy in which correct outputs from a biocomputing solution can produce. For example, in \cite{barros2021engineering}, it is shown how biological noise can affect biocomputing solutions using astrocytes and calcium as the information molecules, and \cite{adonias2020utilizing} shows how neuronal action potentials used to implement logic gate functions are impacted by neuronal noise. Typically, the accuracy, or the number of times these biological logic gates produce correct outputs, is around 90\%. The main goal now is to improve such a number and provide solutions that perform at the optimal accuracy levels when compared to the digital counterpart solutions.


The communication paradigm adopted by neurons, called neuro-spike communication, represents a hybrid molecular communication scheme due to the use of both molecular carriers, depicted by neurotransmitters, and electrical impulses, named action potentials (APs). Specifically, neuro-spike communication can be divided into the phases of axonal transmission, during which APs travel along the axonal channel of the presynaptic neuron, synaptic transmission, which see the transduction from the electrical to chemical stimuli and their propagation to the postsynaptic neuron through chemical synapses, and finally spike generation, that determines the electrical excitation of the postsynaptic neuron \cite{aghababaiyan2017axonal}, \cite{aghababaiyan2019capacity}. In the future may be possible to develop neuronal implantable chips made by biological cells, exploiting neuro-spike communication and synthetic biology techniques \cite{barros2021engineering}. 
Since that action potential can be considered as an "all or nothing" response, the neuron's resting state can be marked to the low logic level, while the excitation state to the high logic level. A neuronal logic gate should be composed of a network of a certain number of neurons, with two neurons in the first layer, which, if correctly stimulated, transmit the input signal to the network, and with one neuron in the last layer which expresses the output. The objective is to design a network able to reproduce the input-output relation of a specific logic gate. 

Our brain is never at rest, at each instant countless neurons generate electrical signals, with different time synchronisation, resulting in unpredictable behaviour. Moreover, each neuron can generate action potentials endogenously, which means in a spontaneous way without any synaptic connections. The summation of all these processes results in enormous background noise, which makes it difficult to understand neurons' signalling. Thinking about biological logic circuits, the use of noisy signals in addition to poor input signals synchronisation could lead to low accuracy in reproducing the desired logic output. Another problem is related to the fact that while in digital logic gates we can use some fixed threshold to define the logic levels ($V_{IL}$, $V_{IH}$, $V_{OL}$, $V_{OH}$), neurons firing threshold is dynamic. The reason for this is because the generation of an action potential is regulated both by the spatial summation and temporal summation mechanisms, and so the excitable threshold is not only related to the resultant input amplitude, but also to its frequency. As a consequence, if the operating frequency does not belong to a certain range, the gate could not follow the expected logic function. For these reasons, a strategy to improve neuronal logic gates' robustness is needed. 


Setting the problem in a more realistic biological environment, our central nervous system is also populated by non-excitable cells, called neuroglia. Astrocytes are a specific type of glial cells, which have the main functions of providing structural support to neurons, supplying them with nutrients and oxygen, insulating them, and protecting them from pathogens through the blood-brain barrier (BBB). Moreover, in recent years it has been demonstrated that astrocytes are in direct communication with neurons \cite{araque1999tripartite,haydon2001glia,fields2002new}. In particular, they can modulate the amount of neurotransmitter released in the synaptic cleft, contributing to the synaptic current regulation and providing feedback to the neuronal activity. Recent experiments demonstrated that astrocyte activity does not simply mirror neuron patterns but seems to produce frequency or amplitude modulation of the neuronal activity \cite{de2009glutamate}. Although the physiologic meaning of astrocytes' information processing remains substantially not clear, their role in neuronal communication seems to be fundamental. These considerations suggest that astrocytes-neuron communication could be exploited in biological logic gates to obtain better accuracy in performing the desired logic function. Tripartite synapses involving engineered astrocytes could be used as a sort of denoising strategy, in order to reduce the background noise that affects synapses. 

In this paper, we propose a computational model of a biological network that exploits neurons and astrocytes to realise simple logic functions. We show how astrocytes contribute positively to the reliability of biocomputing solutions using the electrophysiological activity of mammalian cells. Our contributions include:
\begin{itemize}[\setlength\topsep{0pt}]
    \item \textbf{Modified tripartite synapse activation function supporting spike pattern diversity}: The Izhikevich model \cite{izhikevich2003simple} is adopted for modeling neurons. Tripartite synapses are described using the model presented by Postnov et al. \cite{Postnov2007FunctionalMO}, \cite{postnov2009dynamical} but substituting the synapse activation function with the synaptic conductance model with instantaneous rise and single-exponential decay. By introducing this modification, the astrocyte-neurons coupling can be generalized to different neuronal spike patterns.
    \item \textbf{Biologically plausible model of neuronal AND and OR gates implemented with a neurons-astrocytes network}: A network of two input neurons linked to the same output neuron through a tripartite synapse is used to implement a logic gate. Each cell is simulated using computational models with biological significance. Both OR and AND gating responses are achieved by changing the synaptic strengths and the astrocytes' control parameters.
    \item \textbf{Reliable biocomputing strategy relying on astrocyte-based denoising}: Synaptic Gaussian noise is used to simulate the crosstalk synaptic activity of surrounding neurons. Neuronal AND and OR gates are tested in the presence of noise, both with disabled and enabled astrocytes coupling. The possibility of using the astrocytes' regulation activity to reduce the synaptic noise sources is assessed.
\end{itemize}
Our results demonstrate the effectiveness of astrocytes regulating activity as a denoising mechanism for both AND and OR gates. Analysing both the logic operation error rate (LER) and accuracy, we have observed an improvement in the gating binary responses up to the 25\%. These results pave the way to increased interest in biocomputing technology using mammalian cells, and the future of integrating it with living implants in-vivo for future treatment of neurological diseases. 

\section{Related Works}
Biocomputing has been achieved mostly with genetic or protein circuits before moving towards being implemented in mammalian cells. Cells use signalling proteins for growth, differentiation, migration, or death - meaning proteins are used as carriers of information that encodes cellular functionality \cite{dueber2004rewiring}. Novel signaling functions exploit modulation of proteins' concentration for the generation of advanced functionalities \cite{dueber2003reprogramming}. 
Prehoda et al. \cite{prehoda2000integration} pointed out that operations that can be interpreted as logic gating are already implemented in the biological environment. Specifically, the N-WASP protein, which contributes to the polymerisation during cell motility, manifests its activity only in the case in which two specific ligands bind together its domains. Since individually these ligands are weak activators, i.e. they are not able to elicit the N-WASP activation, but together yield potent activation, this protein circuit acts similarly to an AND gate. 

In \cite{dueber2003reprogramming} Dueber et al. synthetically engineered the N-WASP output domain managing to develop a library of gates, which shows several gating behaviours, such as AND gate and OR gate. In spite of the high design flexibility of protein circuits, they do not result in a simple binary response, but the precise output response depends on the input concentrations \cite{dueber2003reprogramming}. Moreover, if the input concentration changes, the protein gate can behave as a different kind of logic gate. Binary responses of logic gates are desired for the control and stability of computing systems. In addition, it allows the usage of a wide variety of existing binary-based information processing techniques.

More recently, researchers have tried to develop logic gates using neuro-spike communication, intending to achieve different biomedical applications compared to genetic/protein circuits. In a recent work by Adonias et al. \cite{adonias2020utilizing}, they formulated a mathematical model of logic gates made by neurons in a type-rich biological network, and they simulated the effects of its use as a treatment for epileptic seizures. OR gates and AND gates were built with three neurons described by the Hodgkin-Huxley model, using fixed synaptic weights, but changing the probability of establishing a synapse. They demonstrated that increasing firing frequency leads to lower accuracy in performing the specific logic function. Using a more “machine-learning like” approach, but keeping biological plausibility, in \cite{kampakis2012improved} Kampakis et al. presented an artificial neural network architecture, based on the Izhikevich neuron model, able to simulate all logic gate types (AND, OR, NOT, XOR, NOR, NAND, and XNOR). Specifically, they used a genetic algorithm, which changes the weights of connections, and so trains the network to reproduce the desired logic function. Barros et al. \cite{barros2021engineering} proposed an in-vitro realisation of AND gate and OR gate using astrocytes with engineered Ca$^{2+}$ signaling threshold through synthetic genes. They developed a reinforced learning platform, which is able to find the optimal Ca$^{2+}$ activation level and the optimal input transmission period that minimise the noise and the delay. Wet-lab experiments involving engineered human astrocytomas were used to determine astrocytes' sensitivity and validate the reinforced platform results.

From the literature, it is clear that the recent developments in spike-based logic gates realisation were not coupled with the astrocytic role in neuronal communications. For many reasons, one must look closely at how astrocytes regulate the synapses and how this phenomenon can be used for counteracting possible synaptic noise sources. Recent works in neurobiology \cite{araque1999tripartite,haydon2001glia,fields2002new} suggest that astrocytes are the major regulators of neuro-spike information, and we must clarify and develop models that highlight the astrocytes' contributions to the formation of logic gates in neuronal living cells.

\section{Neuro-Astrocyte Logic Gates}

\subsection{Neuro-Astrocyte Model}
\begin{figure*}
    \centering
    \includegraphics[width=0.70\textwidth]{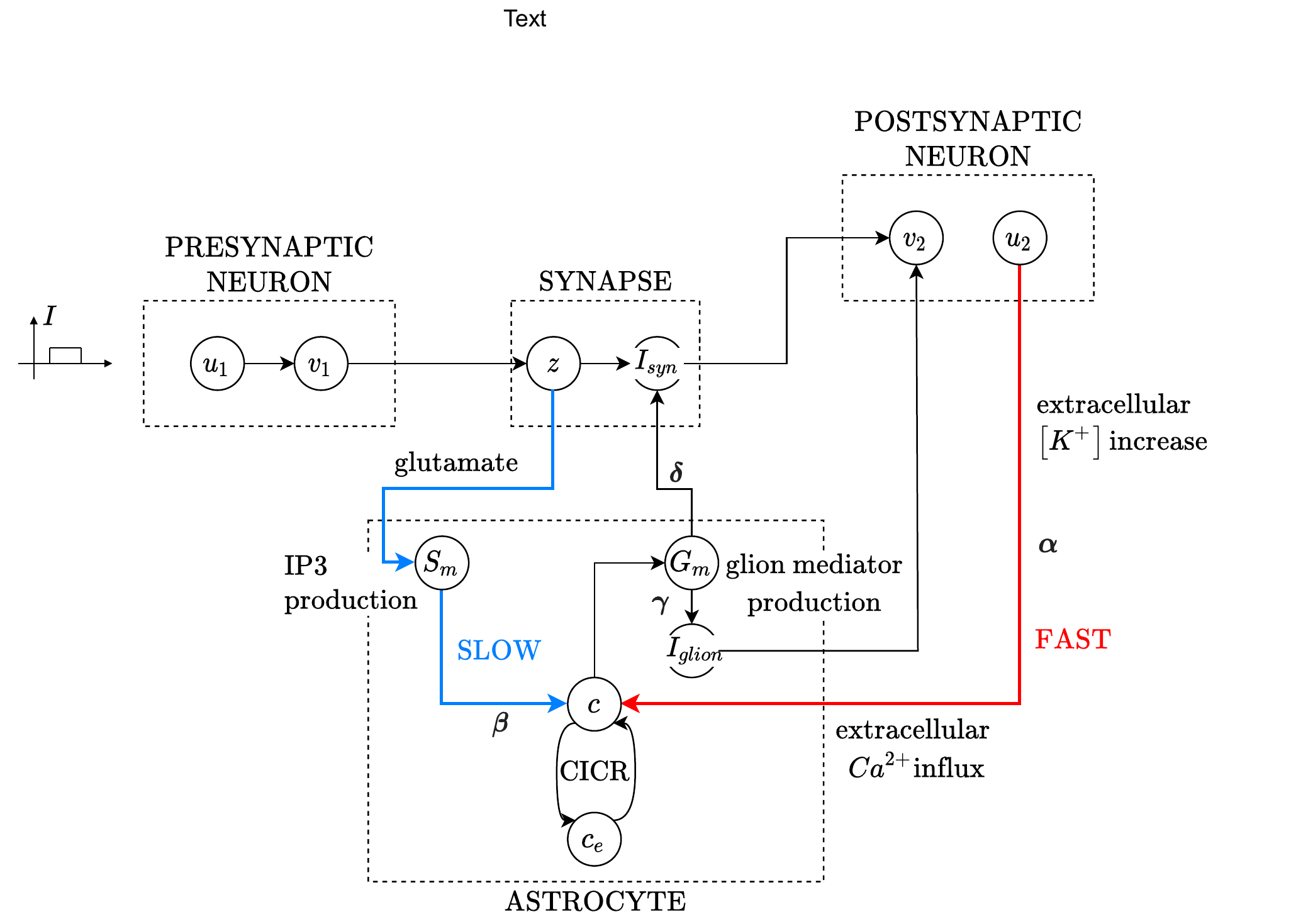}
    \caption{Schematic representation of the tripartite synapse model. The main state variables of each compartment are represented by circles. The arrows describe the dependence relationships. The light blue line shows the astrocyte slow activation pathway, regulated by control parameter $\beta$, whereas the red line indicates the fast activation pathway, regulated by $\alpha$. Control parameters $\gamma$ and $\delta$ influence the astrocyte response.}
    \label{astro1}
    \vspace{-5pt}
\end{figure*}
Accordingly to our purpose, we searched for a flexible spiking model, suitable for coupling with tripartite synapses models, and able to reproduce various spike patterns related to different types of neurons. The Izhikevich model \cite{izhikevich2003simple} provides an accurate description of a spiking neuron and we explore its coupling with the Postnov et al. model of tripartite synapse \cite{Postnov2007FunctionalMO}, which both are described in the following. We improve upon this model by developing a solution to couple both models for many spike patterns.

The Izhikevich model \cite{izhikevich2003simple} achieves an accurate description of spike patterns through the following equations:
\begin{equation} \label{Izhi_1}
    \frac{dv}{dt}=0.04v^2+5v+140-u+I \,,
\end{equation}
\begin{equation} \label{Izhi_2}
    \frac{du}{dt}=a(bv-u) \,,
\end{equation}
with the auxiliary after-spike resetting:
\begin{equation} \label{Izhi_cond}
    \text{if} \quad v\geq+30 \text{ mV,} \quad \text{then} \begin{cases} v	\leftarrow c \\ u 	\leftarrow u+d \,. \end{cases}
\end{equation}
\noindent where $v$ and $u$ are the state variables. The variable $v$ represents the membrane voltage potential of the neuron. The variable $u$, called the recovery variable, accounts for the activation of K$^+$ ionic currents, and the inactivation of Na$^+$ ionic currents, and provides negative feedback to $v$. $I$ represents the stimulating current. When the condition defined in Eq. \ref{Izhi_cond} is verified, the neuron fires, and $v$ and $u$ are reset. $v$ is first set to 30 and then to $c$, in order to have all spikes with the same amplitudes \cite{izhikevich2004model}. The parameters $a$,$b$,$c$, and $d$ are used to define a specific neuronal spike pattern. Not all combinations of these parameters result in a physiological behaviour, but Izhikevich defined 20 default combinations that manage to simulate the most important spike patterns.

As previously mentioned, neurons' activity is heavily coupled with astrocytes activity, which regulates the amount of neurotransmitter released in the synaptic cleft. Astrocytes are not able to generate APs, but instead, they encode information through oscillations in Ca$^{2+}$ concentration. Considering a tripartite synapse, namely a synapse involving two neurons and an astrocyte, when the presynaptic neuron fires it induces the release of neurotransmitter, such as glutamate, in the synaptic cleft. The glutamate binding to astrocyte receptors triggers astrocyte activity, by increasing 1,4,5-trisphosphate (IP3) intracellular concentration \cite{de2009glutamate,haghiri2014investigation}. As a consequence, IP3 binds the receptors placed on the astrocyte endoplasmic reticulum (ER), and Ca$^{2+}$ is realised from ER to the cytoplasm. Since the opening of IP3 channels increases with increasing Ca$^{2+}$ concentrations, this leads to a self-amplifying release mechanism, called calcium-induced calcium release (CICR). With high Ca$^{2+}$ concentrations, the astrocyte responds by releasing "glion mediators" (or "glion transmitters"), such as ATP and glutamate, in the synaptic cleft. This contributes to the synaptic transmission \cite{di2007calcium} enabling the feedback mechanism. Moreover, this could be positive or negative feedback, depending on if the astrocyte glutamate release happens near an excitatory or inhibitory interneuron \cite{nadkarni2003spontaneous}. Although several mediators are involved in the neuron-glion interaction, we can define two main pathways of astrocyte activation \cite{Postnov2007FunctionalMO}: the slow activation pathway and the fast activation pathway. The previously explained mechanism of activation via IP3 production is defined as the slow activation pathway. In addition, astrocytes can be activated due to the increase of K$^+$ extracellular concentration. Since this kind of astrocyte activation can be considered instantaneous, this mechanism is called the fast activation pathway.

We consider the tripartite synapse model defined by Postnov et al. \cite{Postnov2007FunctionalMO} coupled with the Izhikevich model, as proposed in \cite{haghiri2014investigation}. Defining the state variables $z$ as the synaptic activation variable \cite{kopell2000gamma}, $I_{syn}$ as the synaptic current, and $I_{glion}$ as the astrocyte-induced current, the synaptic coupling is described by the following equations:
\begin{equation} \label{ODE z}
 \tau_s \frac{dz}{dt}=[1+tanh(s_s(v_1-h_s))](1-z)-\frac{z}{d_s} \,,
\end{equation}
\begin{equation}
    I_{syn}=(k_s-\delta G_m)(z-z_0) \,,
\end{equation}
\begin{equation} \label{I glion}
    I_{glion}=\gamma G_m \,,
\end{equation}
\noindent where $v_1$ is the presynaptic neuron potential and $\tau_s$ describes the time delay. Parameters $h_s$, $s_s$, $d_s$ control the activation and relaxation of $z$, $k_s$ is the conductance and $G_m$ is the glion mediator concentration. Control parameters $\delta$ and $\gamma$ regulate the strength of the astrocyte influence on synapse and postsynaptic neuron respectively. $z_0$ represents the reference level for $z$ (e.g. when the presynaptic neuron is silent, $z(t)=z_0$). If $v_1<h_s$ the synapse is inactive ($z=0$), while when $v_1-h_s$ term is positive the synapse is active, and $z$ generates the synaptic current $I_{syn}$. Both $I_{syn}$ and $I_{glion}$ are used as stimulating currents in the Izhikevich model of the postsynaptic neuron. The dynamics of calcium concentration $c$ within the astrocyte and calcium concentration $c_e$ in the ER are described by the following two-pool model modified with two additional
terms concerning the external forcing via the tripartite synapse:
\begin{equation} \label{ODE c}
    \tau_c\frac{dc}{dt}=-c-k_4f(c,c_e)+(r+\alpha u_2+\beta S_m) \,,
\end{equation}
\begin{equation}
    \varepsilon_c \tau_c \frac{dc_e}{dt}=f(c,c_e) \,,
\end{equation}
with
\begin{equation}
   f(c,c_e)=k_1 \frac{c^2}{1+c^2}-\left( \frac{c_e^2}{1+c_e^2} \right) \left( \frac{c^4}{k_2^4+c^4} \right) -k_3c_e \,,
\end{equation}
where $\tau_c$ defines the time scale for the calcium oscillations with time separation parameter $\varepsilon_c$ and $k_1$, $k_2$, $k_3$, and $k_4$ are constant parameters. In Eq. \ref{ODE c}, the term $-c$ represents the contribution of Ca$^{2+}$ pumps, which pump ions out of the cell. The term $r$ is a steady flux of Ca$^{2+}$ into the cell without external influence (when $\alpha=0$, $\beta=0$). The external forcing term $\alpha u_2$, with $u_2$ postsynaptic neuron recovery variable, represents the calcium influx from the extracellular space due to glion depolarisation (fast activation pathway). The external forcing term $\beta S_m$ is the influx sensitive to synapse mediator $S_m$ production (slow activation pathway). The function $f(c,c_e)$ is a nonlinear function that describes the Ca$^{2+}$ fluxes between the cytoplasm and the ER \cite{mesin2017book}. Its first term, often called uptake, is a Hill-type law (i.e. a sigmoidal-like curve which saturates to a reference level) indicating the increasing of ER calcium concentration depending on cytoplasmatic concentration, with the aim of calcium storage in the ER. The second term, called release, indicates the release of calcium from ER reticulum to the cytoplasm: its second factor is a Hill-type law related to the CICR positive feedback and the first factor is another Hill-type law which represents the flux dependence in the ER calcium concentration (i.e. if there is no calcium in it, nothing can be released, while in the opposite case if there is a lot of calcium the flux saturates at its maximum value). Finally, the third term $-k_3c_e$ represents the loss of ER calcium by leakage channels. Furthermore, this system is coupled with the description of glion mediator production $G_m$ and IP3 mediator $S_m$:
\begin{equation} \label{ODE Sm}
    \tau_{S_m} \frac{dS_m}{dt}=[1+tanh(s_{S_m}(z-h_{S_m}))](1-S_m)-\frac{S_m}{d_{S_m}} \,,
\end{equation}
\begin{align}
    \tau_{G_m} \frac{dG_m}{dt}=[1+tanh(s_{G_m}(c-h_{G_m}))](1-G_m)-\frac{G_m}{d_{G_m}} \,, \notag \\
\end{align}
\noindent where $S_m$ production is triggered by increasing $z$ (similarly as in Eq. \ref{ODE z}), while $G_m$ production is triggered by increasing calcium cytoplasm concentration $c$.
By changing control parameters $\alpha$, $\beta$, $\gamma$, $\delta$ several neuron-astrocyte dynamics can be simulated \cite{postnov2009dynamical}. Parameters $\alpha$ and $\beta$ regulate fast and slow activation pathways respectively. While the fast mechanism produces a fast but short-term response, occurring as a single calcium spike, the slow mechanism elicits long-term activity, since that by increasing $\beta$ the number of calcium spikes increases. Control parameters $\gamma$ and $\delta$ allow controlling the astrocyte feedback performed on the postsynaptic neuron. The parameter $\gamma$ regulates the depolarising current $I_{glion}$, thus it controls the positive feedback mechanism. This can support the postsynaptic firing activity also after the end of the stimulus. The parameter $\delta$ influences the negative contribution subtracted to the synaptic strength $k_s$, thus representing negative feedback that can inhibit the transmission of the stimulus. An overall representation of the tripartite synapse model is shown in Fig. \ref{astro1}

\subsection{Synapse Activation Function} \label{Modifications}
 One of the most prominent features of the Izhikevich model is its ability to accurately reproduce different neuron spike patterns, such as tonic/phasic spike and burst. Instead, our preliminary experimental tests showed that the Postnov model, coupled with Izhikevich neurons, manages to simulate only the most common type of spike pattern, which is the tonic spike. Hence, we propose the following modifications to extend the model also to other spike patterns. Moreover, we expand the model for the case of multiple tripartite synapses, which means more than one presynaptic neuron and one astrocyte connected with the same postsynaptic neuron.
Instead of using the $z$ activation variable, as reported in Eq. \ref{ODE z}, we used the well-known synaptic conductance model with instantaneous rise and single-exponential decay \cite{roth2009modeling}. The synaptic conductance $g_{syn_i}$ between the $i$-th presynaptic neuron and the only one postsynaptic neuron is expressed as:

\begin{equation}
     \begin{gathered}
        g_{syn_i} \leftarrow g_{syn_i} + flag_i \,, \\ 
        \text{with } flag_i=\begin{cases} 1 \text{ if the $i$-th presynaptic neuron is firing} \\ 0 \text{ otherwise} \,, \end{cases}
     \end{gathered}
\end{equation}
\begin{equation} \label{g_2}
    \frac{dg_{syn_i}}{dt}=-\frac{g_{syn_i}}{\tau_g} \,,
\end{equation}
\noindent where $\tau_g=10$ ms and $flag_i$ indicates if the $i$-th presynaptic neuron is firing or not. When a presynaptic neuron fires an AP, the corresponding synapse is activated by the instantaneous increase of $g_{syn_i}$, and then the conductance is extinguished with exponential decay. 
Since the variable $z$ is here substituted by $g_{syn_i}$, also Eq. \ref{ODE Sm} describing the IP3 concentration of the $i$-th astrocyte needs to be modified as follows:
\begin{align}
    \tau_{S_m} \frac{dS_{m_i}}{dt}=[1+tanh(s_{S_m}(g_{syn_i}-h_{S_m}))](1-S_{m_i})-\frac{S_{m_i}}{d_{S_m}} \,. \notag \\
\end{align}

Finally, the resulting synaptic current is formulated using a conductance-based model, with the supplementary contribution accounting for the astrocyte feedback:
\begin{equation} \label{I_syn new}
    I_{syn}=\sum_{i=1}^{n} \left( w_i g_{syn_i}(E_{syn_i}-v_{post}) -\delta_i G_{m_i} \right) \,,
\end{equation}
\noindent where $w_i$ and $E_{syn_i}$ represent the weight and the reversal potential of the $i$-th synapse, $v_{post}$ is the postsynaptic membrane potential and $\delta_i G_{m_i}$ is the $i$-th astrocyte negative feedback contribution. Since all the synapses involved in our model are excitatory, we set $E_{syn_i}=0$. In addition, the postsynaptic neuron is also stimulated by the astrocyte mediator through $I_{glion_i}$ as expressed in Eq. \ref{I glion}. 

\subsection{Logic Gates}
\begin{figure*}
    \centering
    \includegraphics[width=0.70\textwidth]{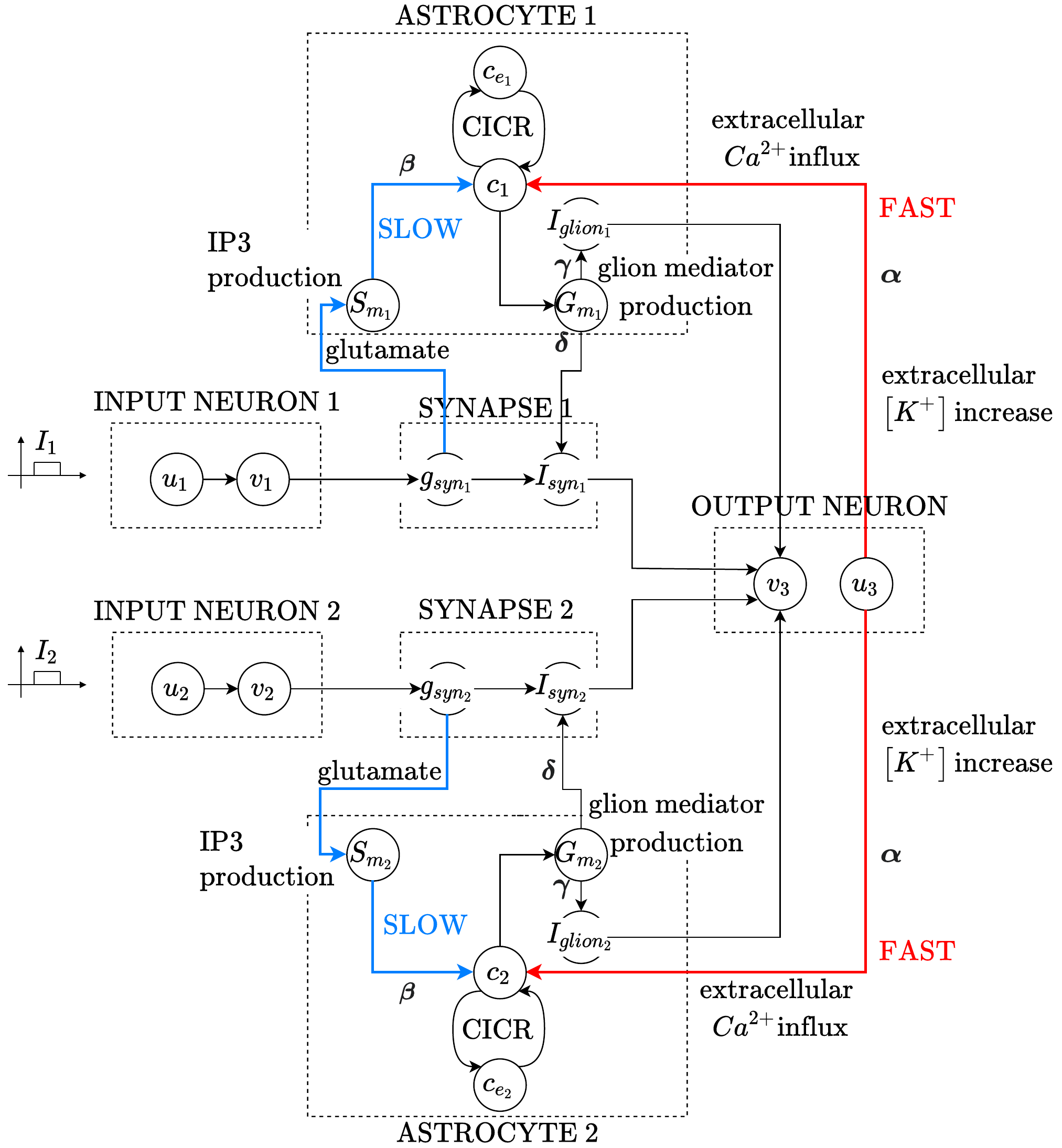}
    \caption{Schematic representation of the logic gate model involving two tripartite synapses. Each neuron and astrocyte is described by its state variables, and the arrows indicate the dependence relationships between variables. For both astrocytes, the slow activation pathway is highlighted with a light blue line, while the fast activation pathway with a red line. The control parameters that describe these processes are reported in bold type close to the lines. In this final network, the postsynaptic membrane potential $v_3$ is influenced by the stimuli received by two distinct tripartite synapses.}
    \label{astro2}
    \vspace{-5pt}
\end{figure*}

In this subsection, we describe the cellular network structure, which is the same in all logic gates developed. As can be observed in Fig. \ref{astro2}, it consists of two presynaptic neurons, called INPUT NEURONS and labelled with indexes 1 and 2, and a postsynaptic neuron, called OUTPUT NEURON and labelled with the index 3. The presynaptic membrane potentials $v_1$ and $v_2$ represent the logic inputs, while the postsynaptic membrane potential $v_3$ is the logic output. $u_1$, $u_2$, and $u_3$ are the neurons’ recovery variables. The presynaptic neurons are connected to the postsynaptic neuron through two distinct synapses, each of them regulated by an astrocyte. $g_{syn_1}$ and $g_{syn_2}$ indicate the synaptic conductance of SYNAPSE 1 and 2. For ASTROCYTE 1 and ASTROCYTE 2 respectively, the variables $c_{e_1}$, $c_{e_2}$ are the Ca$^{2+}$ concentrations in the ER, $c_1$, $c_2$ are the Ca$^{2+}$ in the cytoplasm, $S_{m_1}$, $S_{m_2}$ are the IP3 mediators, $I_{glion_1}$, $I_{glion_2}$ are the astrocyte-induced currents, and finally $G_{m_1}$, $G_{m_2}$ are the glion mediators. The presynaptic neurons are described using Eqs. \ref{Izhi_1} \ref{Izhi_2} \ref{Izhi_cond}. For both of them, the stimulating currents $I$ in Eq. \ref{Izhi_1} are chosen as rectangular functions $I_1$ and $I_2$, which define when the logic inputs are ON or OFF. The astrocytes and the corresponding synapses models are adapted as explained in Subsection \ref{Modifications}. The synaptic current, given by equation \ref{I_syn new}, consists of the sum of the contributions $I_{syn_1}$ and $I_{syn_2}$ coming from each synapse. The overall current that stimulates the postsynaptic neuron is given by the sum of the synaptic current and the currents coming from the two astrocytes:
\begin{equation}
    I_{tot}=I_{syn_1}+I_{syn_2}+I_{glion_1}+I_{glion_2} \,.
\end{equation}
\noindent Therefore the postsynaptic neuron Izhikevich model can be formulated using $I_{tot}$ as the stimulating current:
\begin{equation}
    \frac{dv_3}{dt}=0.04v_3^2+5v_3+140-u_3+I_{tot} \,,
\end{equation}
\begin{equation}
    \frac{du_3}{dt}=a(bv_3-u_3) \,,
\end{equation}
\begin{equation}
    \text{if} \quad v_3\geq+30 \text{ mV,} \quad \text{then} \begin{cases} v_3	\leftarrow c \\ u_3 	\leftarrow u_3+d \,, \end{cases}
\end{equation}
where $u_3$ is the recovery variable of the postsynaptic neuron.

In order to design the OR gate, the synaptic strength $w_i$ and the feedback contributions coming from the astrocytes are set in a such way that the firing of one presynaptic neuron must produce a sufficiently strong stimulation such that it causes the firing of the postsynaptic neuron. Therefore, if we set ON one of the two currents $I_1$ and $I_2$, while the other is OFF, the output must be at the high level. Then to design the AND gate, the basic idea is that we need to make inputs less influential, such that both two input neurons have to be at the high level to make the level of the output neuron high too. This can be obtained by decreasing the synaptic strength, or by regulating the astrocytes' negative feedback. Table \ref{table1} reports the model parameters in common between all logic gates implementations.
 \begin{table}
 \caption{Parameters of the tripartite synapse model in common between all logic gate implementations. All the parameters are dimensionless.}
    \begin{center}
        \begin{tabular}{c | c | c | c | c}
        \hline
        Variable & Value & & Variable & Value \\
        \hline
        \hline
        $k_1$ & $0.13$ & & $k_2$ & $0.9$ \\ 
        $k_3$ & $0.004$ & & $k_4 $ & $2/\varepsilon_c$ \\
        
         $\varepsilon_c$ & $0.04$ & &
         $r$ & $0.31$  \\
        
          $\tau_c$ & $8$ & & $\tau_{S_m}$ & $100$ \\ 
          $\tau_{G_m}$ & $50$ & & 
          $\tau_{g}$ & $10$ \\
        
         $s_{S_m}$ & $100$ & & $s_{G_m}$ & $100$  \\
        
         $h_{S_m}$ & $0.45$ & & $h_{G_m}$ & $0.5$ \\
        
         $d_{S_m}$ & $3$ & & $d_{G_m}$ & $3$ \\
        \hline
        \end{tabular}
        \label{table1}
    \end{center}
    \vspace{-10pt}
    \end{table}

\subsection{Noise Model} \label{noise model}
\begin{figure*}[htp]
\begin{minipage}[t]{.49\textwidth}
\centering
\subfloat[inputs {[1 0]}]{\includegraphics[width=.47\linewidth]{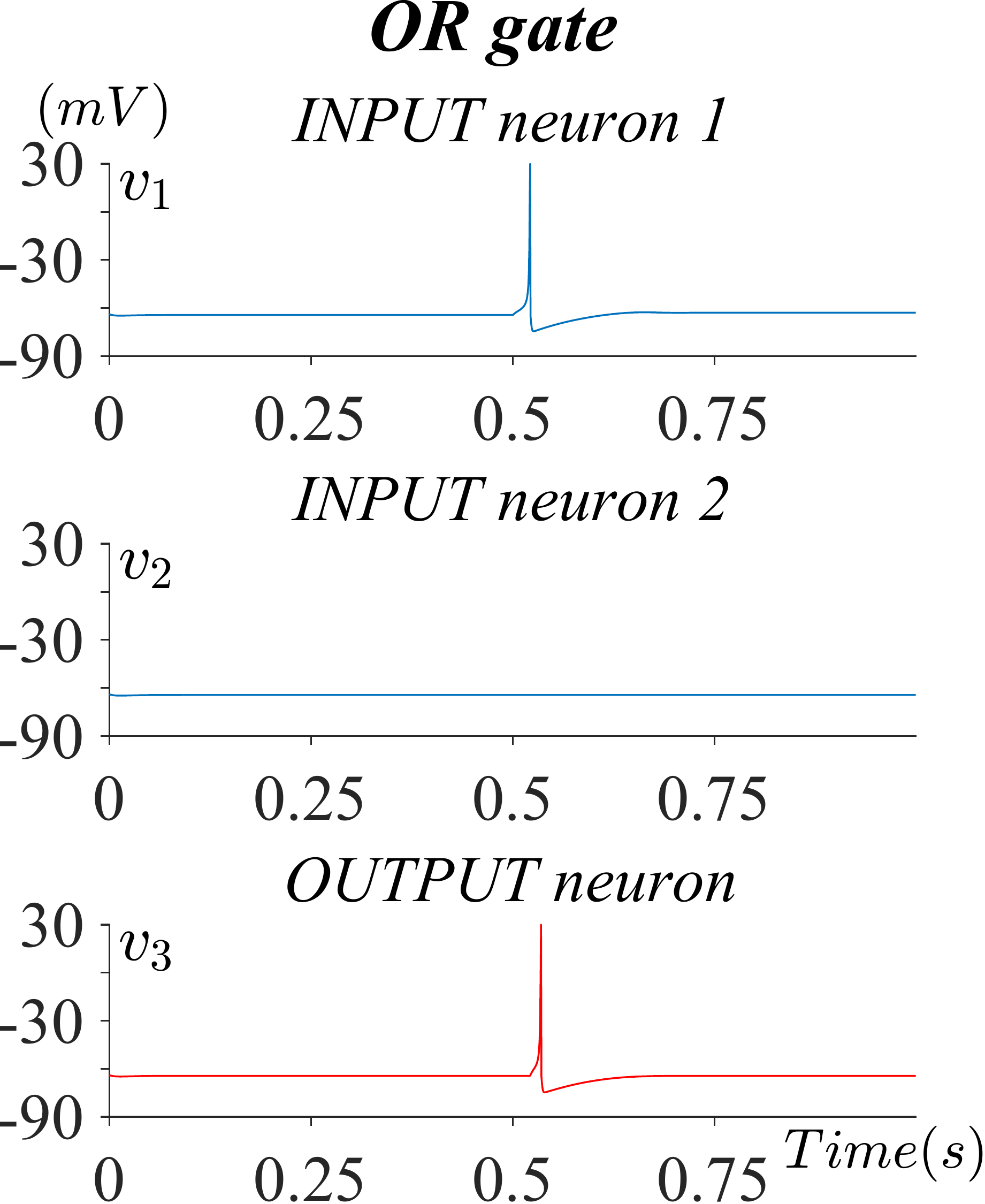}}\quad
\subfloat[inputs {[1 1]}]{\includegraphics[width=.47\linewidth]{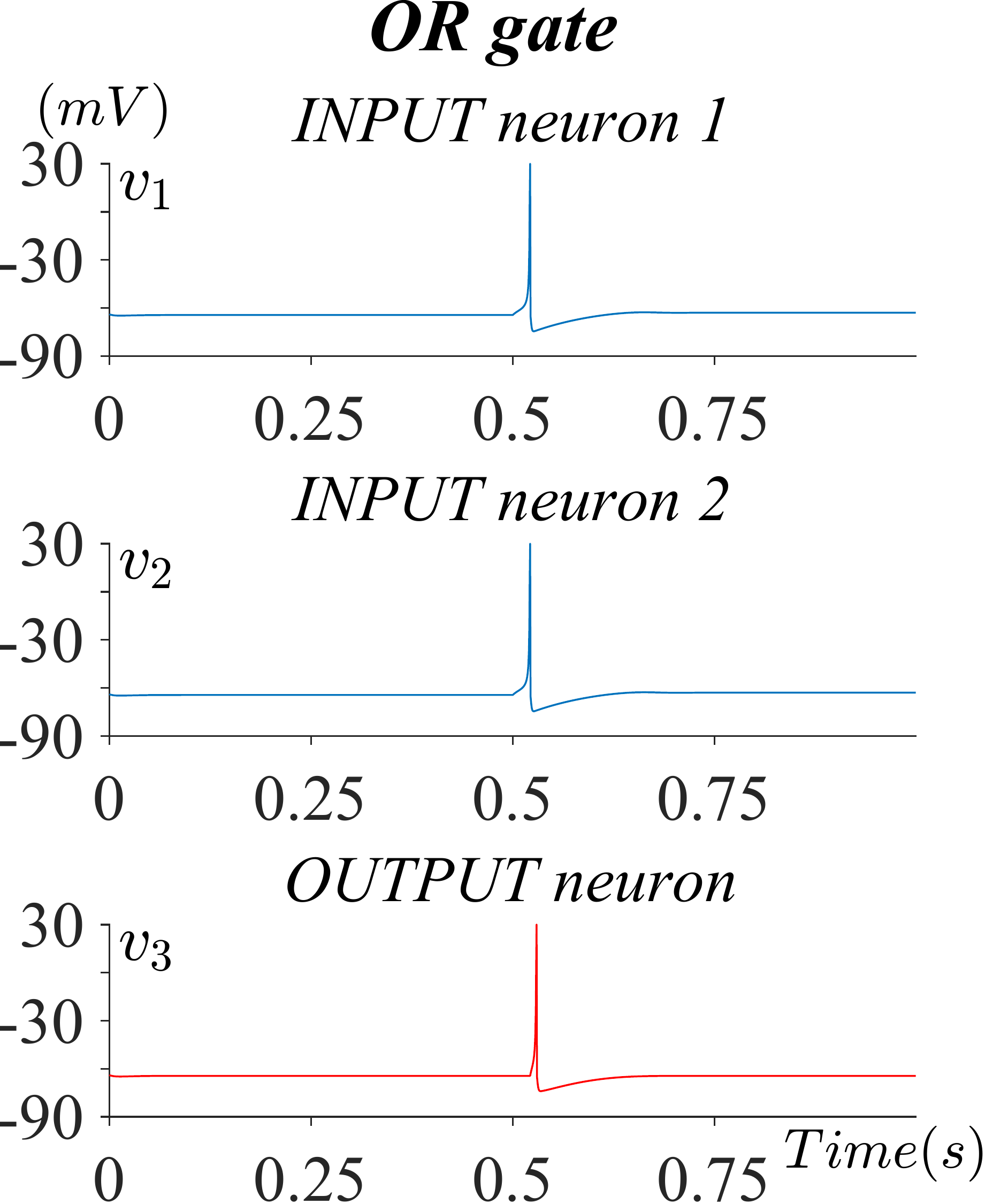}}
\caption{Phasic pattern OR gate involving only neurons, with stimulating current $I=0.5$ pA and synaptic strength $w_{i}=0.02$.}
\label{phasic1}
\end{minipage}\hfill
\begin{minipage}[t]{.49\textwidth}
\centering
\subfloat[inputs {[1 0]}]{\includegraphics[width=.47\linewidth]{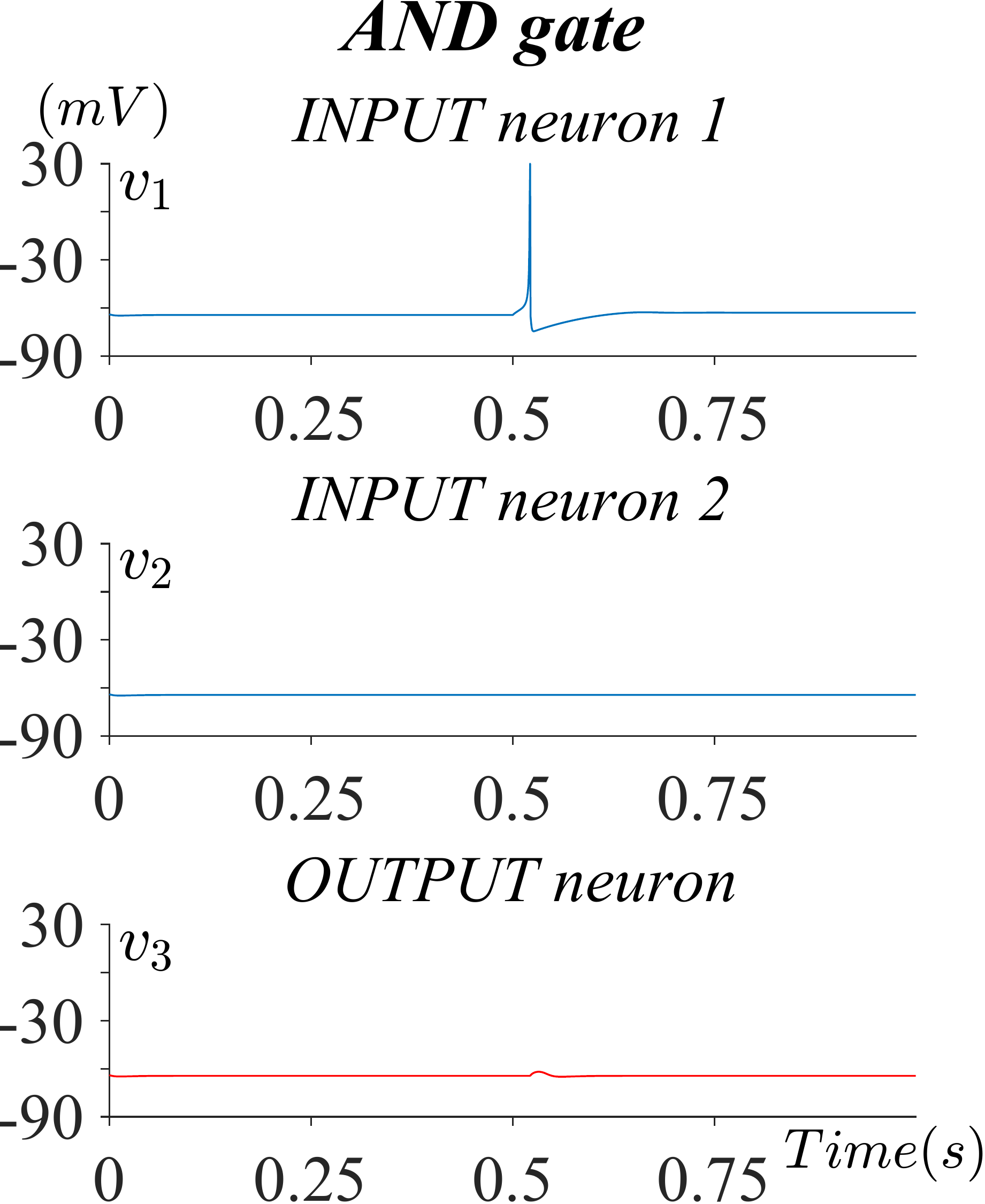}}\quad
\subfloat[inputs {[1 1]}]{\includegraphics[width=.47\linewidth]{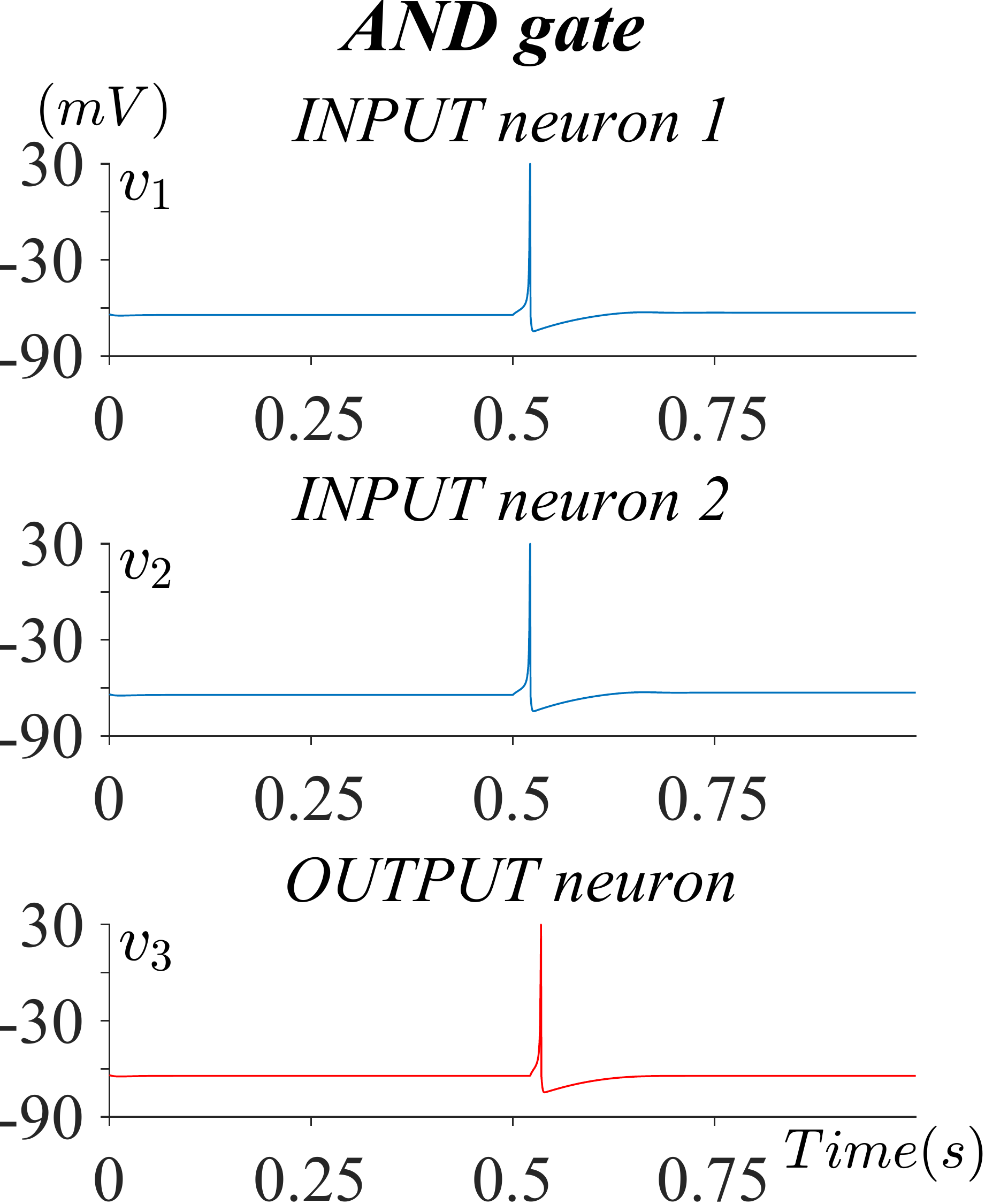}}
\caption{Phasic pattern AND gate involving only neurons, with stimulating current $I=0.5$ pA and synaptic strength $w_{i}=0.01$.}
\label{phasic2}
\end{minipage}\hfill
\begin{minipage}[t]{.49\textwidth}
\centering
\subfloat[inputs {[1 0]}]{\includegraphics[width=.47\linewidth]{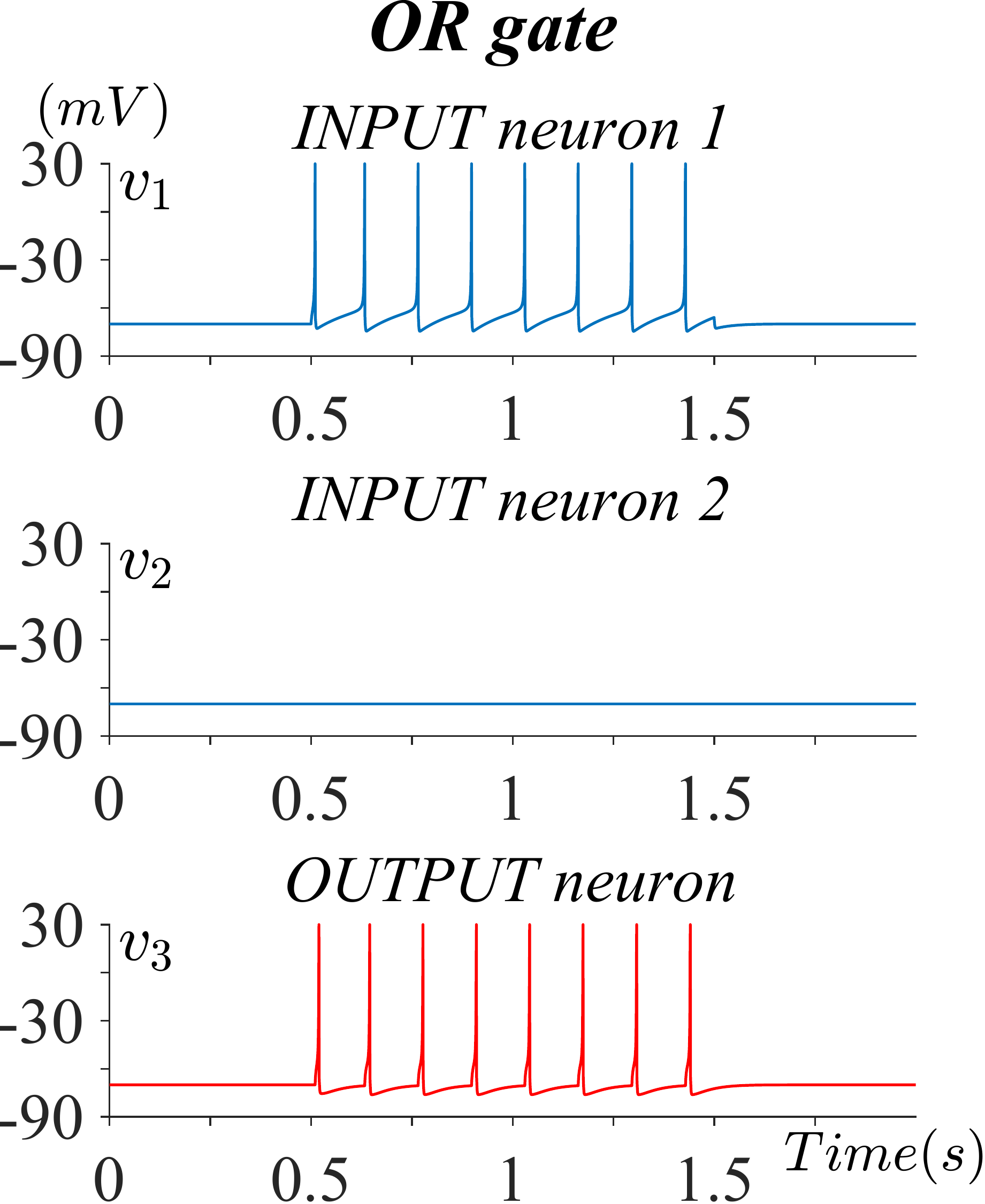}}\quad
\subfloat[inputs {[1 1]}]{\includegraphics[width=.47\linewidth]{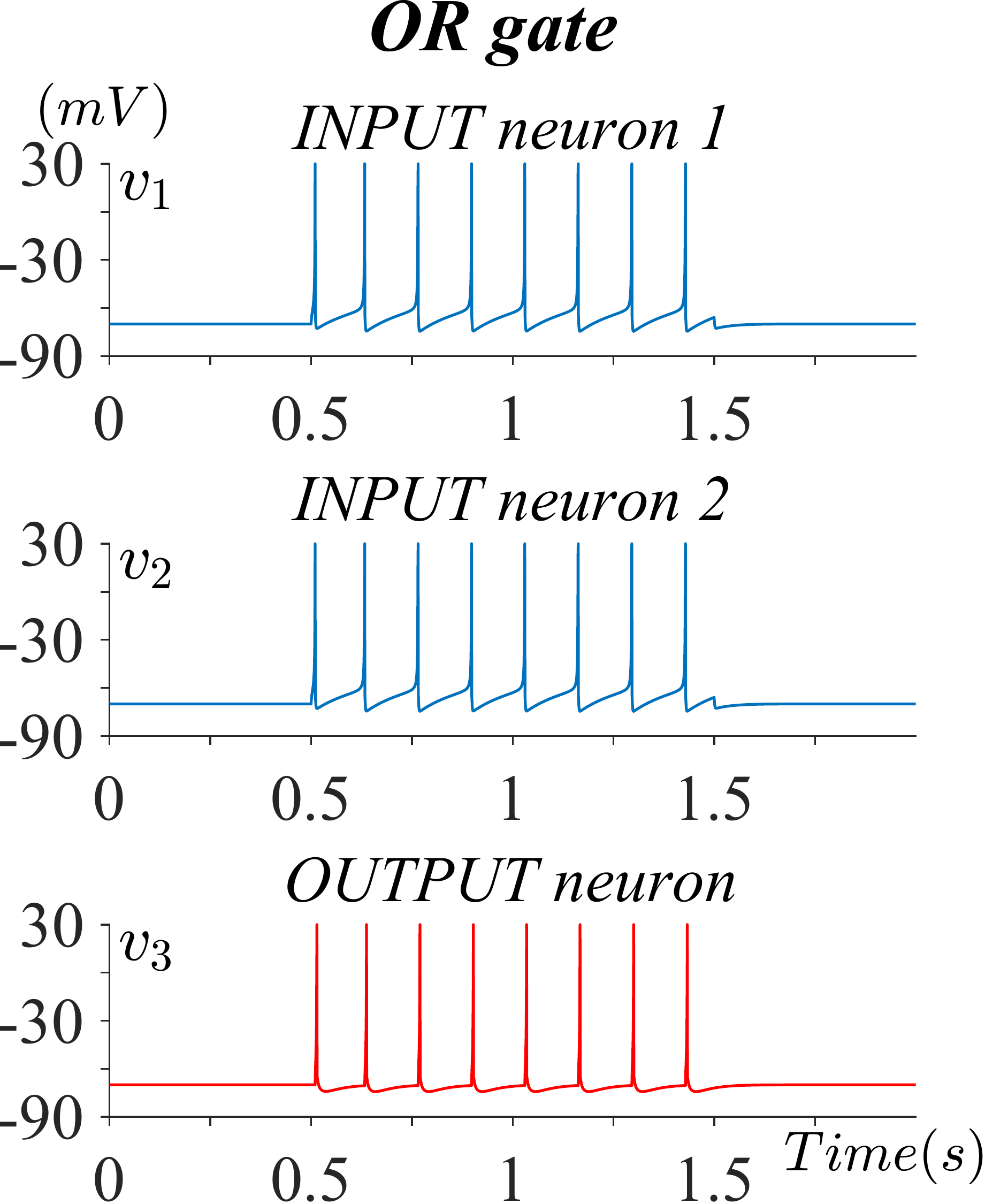}}
\caption{Tonic pattern OR gate involving only neurons, with stimulating current $I=4$ pA and synaptic strength $w_{i}=0.09$.}
\label{tonic1}
\end{minipage}\hfill
\begin{minipage}[t]{.49\textwidth}
\centering
\subfloat[inputs {[1 0]}]{\includegraphics[width=.47\linewidth]{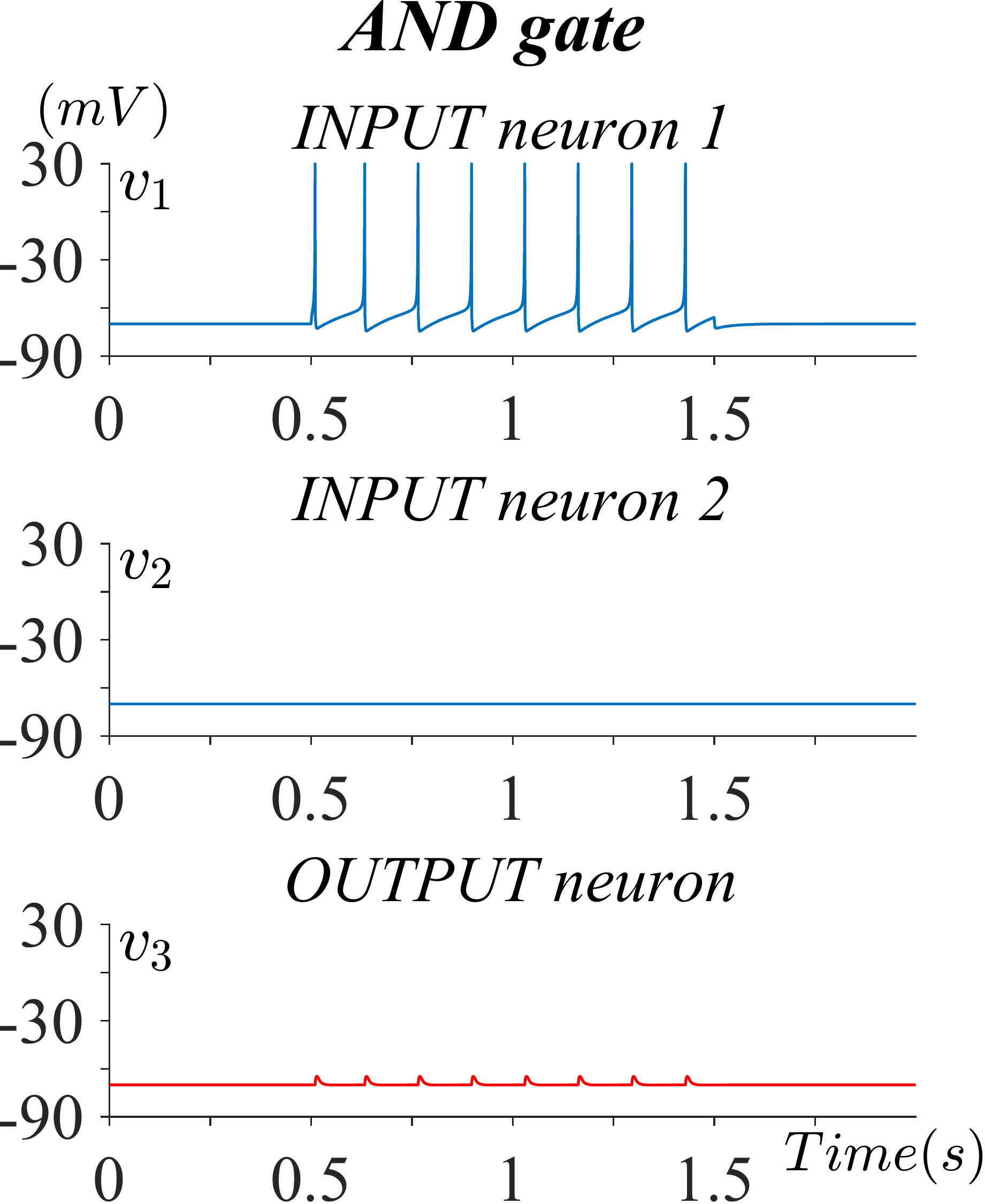}}\quad
\subfloat[inputs {[1 1]}]{\includegraphics[width=.47\linewidth]{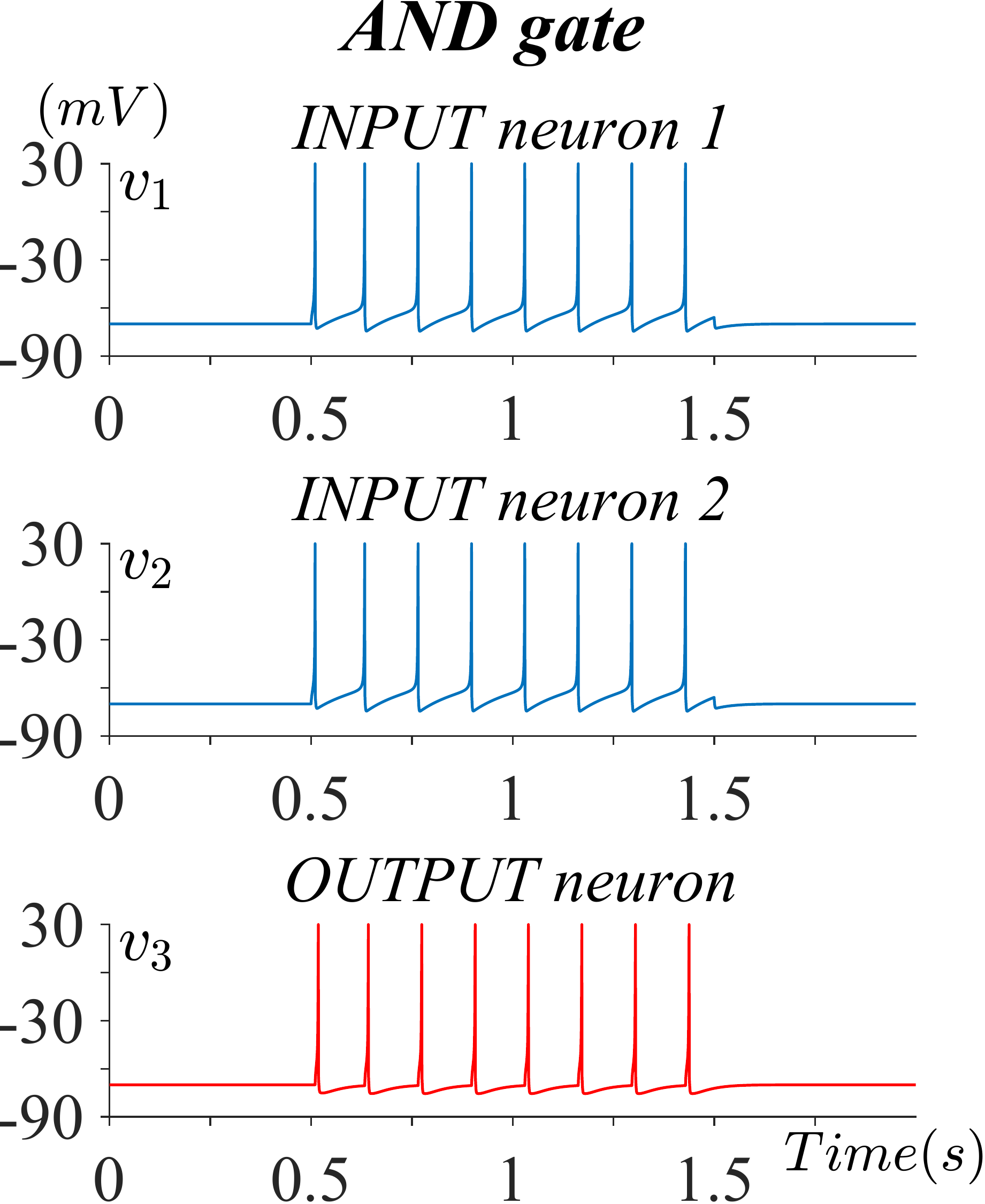}}
\caption{Tonic pattern AND gate involving only neurons, with stimulating current $I=4$ pA and synaptic strength $w_{i}=0.05$.}
\label{tonic2}
\end{minipage}
\vspace{-5pt}
\end{figure*}
We want to explore the effects of noise on the information encoded in spikes that ultimately can affect the output of the logic gates. This approach also allows us to develop an improved verification of our models, and investigate how astrocytes can handle noise through its feedback mechanisms.
In physiological networks, each synapse is influenced by the activity of thousands of neighbouring synapses, resulting in a background synaptic noise. With the simplifying assumption of sources as independent and identically distributed random variables, the sum of these noisy synaptic effects converges to a Gaussian probability distribution, in accordance with the central limit theorem. For us, noise represents an arbitrary effect on the information of the spikes, but is not focused on particular parts of the brain or specific scenarios that the presented noise model may need to be readjusted.
Synaptic noise is simulated using a normal distributed random variable \cite{brunel2001effects}, with zero mean and variance $\sigma^2$, which contributes to the stimulating current of the postsynaptic neuron:
\begin{equation}
    I_{noise}\sim N(0,\sigma^2) \,,
\end{equation}
\begin{equation}
    I_{tot}=I_{syn_1}+I_{syn_2}+I_{glion_1}+I_{glion_2}+I_{noise} \,.
\end{equation}

\section{Results}
All the simulations here reported were implemented in MATLAB R2019b, and the code is available at \cite{bassogiulio_2022_6890378}. The equations of the model were solved numerically using the explicit Euler method with step size $dt=0.5$ ms.

\subsection{Logic Gates with Different Spike Patterns} \label{Results A}
First of all the logic gates are tested using only neurons and disabling astrocytes activity. This can be done by setting astrocytes control parameters $\alpha$, $\beta$, $\gamma$, $\delta$ to zero. With this step we focus only on the logic gating, meaning that we want to understand how a set of inputs represented by spike patterns can be transformed into an output spike pattern implementing a Boolean function. AND gate and OR gate are developed investigating two possible spike patterns, specifically phasic spike and tonic spike pattern. 

Phasic spike occurs when a neuron fires a single spike at the onset of the stimuli and then remains at the resting state \cite{izhikevich2004model}. In the first realisation, all the three neurons in the logic gate network communicate using phasic spike. This can be done by setting the initial values of all the voltage potentials $v_i$, the recovery variables $u_i$, and the values of the model parameters $a$,$b$,$c$,$d$ as indicated in \cite{izhikevich2003simple}. The stimulating current of both presynaptic neurons is chosen as a rectangular function, with amplitude equal to $I=0.5$ pA between 0.5 and 1.5 seconds and null otherwise. Then OR gate and AND gate are developed regulating the influence of the stimuli, through the synaptic strength $w_i$. For this purpose, both OR gate synaptic strengths are chosen as $w_i=0.02$. With this choice, one high input is enough to stimulate an output response. For the AND gate, the synaptic strengths are chosen as $w_i=0.01$, in a such way that both inputs need to be at the high level to elicit a response. Figures \ref{phasic1} and \ref{phasic2} report two examples of logic gate operation with phasic spike pattern.

With tonic spike pattern, the neuron continues to fire as long as the stimulating current is ON. The combination of neuron model parameters that allows to simulate tonic spike pattern can be found in \cite{izhikevich2003simple}. When the stimulating current is ON, its amplitude is chosen as $I=4$ pA. The OR gate is realized using synaptic weights $w_i=0.09$, while for the AND gate they are set as $w_i=0.05$. Figures \ref{tonic1} and \ref{tonic2} show two examples of logic gate dynamics using tonic spike. Here each spike, or absence of spike, can be considered as a bit of a binary signal. 

\subsection{Logic Gates using Astrocytes}
\begin{figure}
\centering
  \begin{subfigure}{0.49\columnwidth}
    \includegraphics[width=\linewidth]{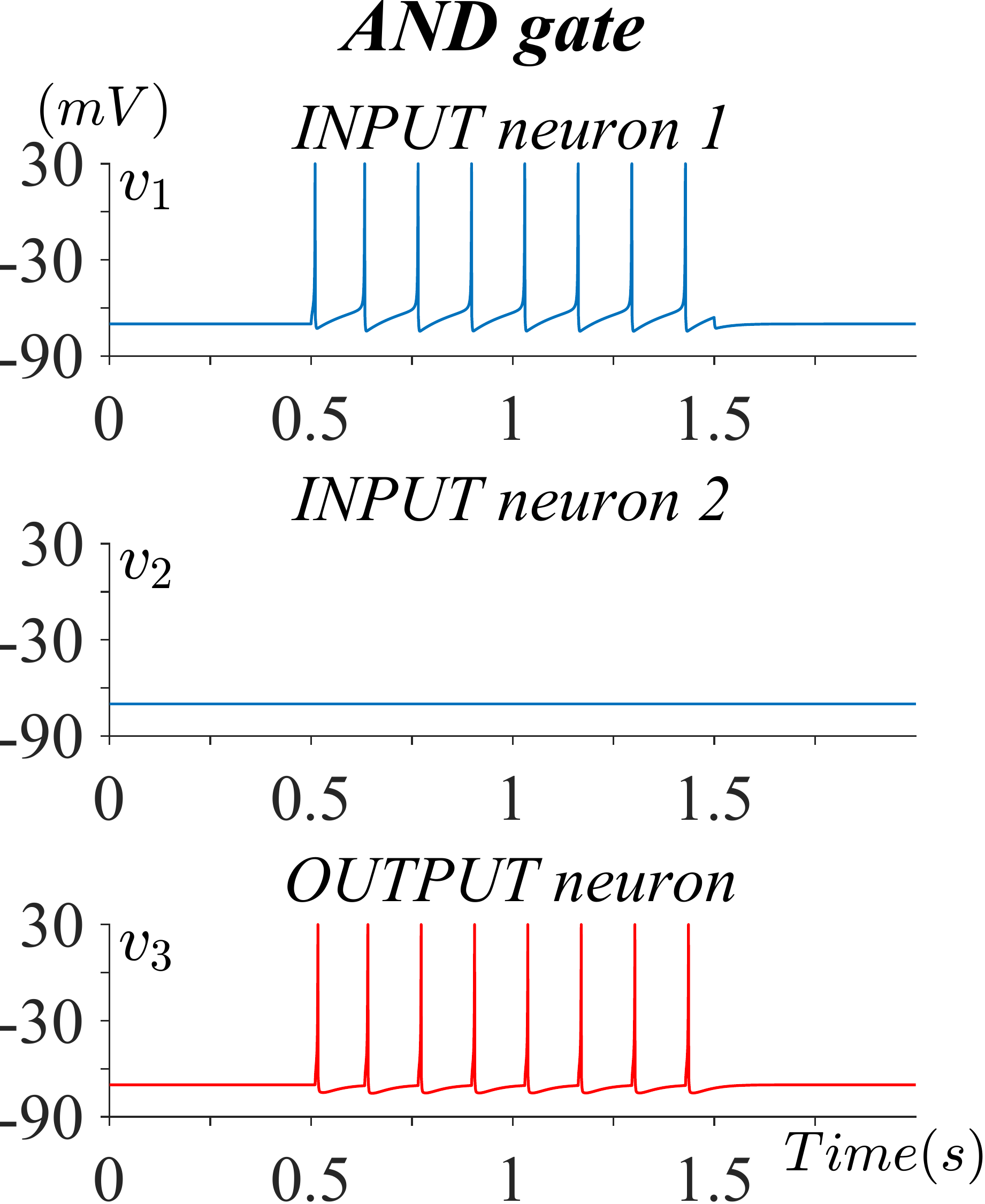}
    \caption{inputs [1 0]}
    \label{B1-2a}
  \end{subfigure}
  \hfill 
  \begin{subfigure}{0.49\columnwidth}
    \includegraphics[width=\linewidth]{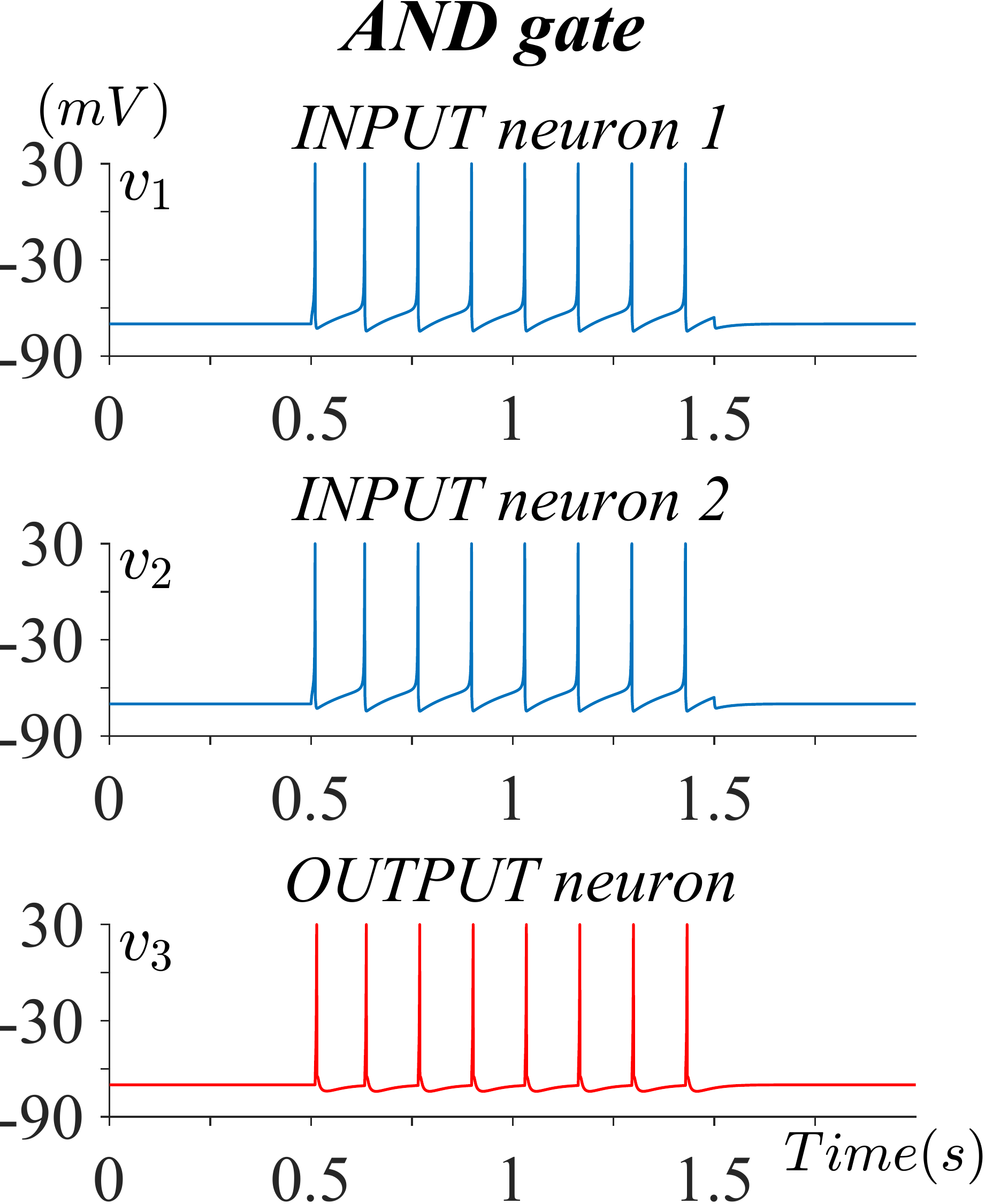}
    \caption{inputs [1 1]}
    \label{fig:2}
  \end{subfigure}
  \caption{Mismatched AND gating due to the too high synaptic strength ($w_i=0.11$). The logic gate is obtained involving only neurons and with stimulating current $I=4$ pA.}
  \label{B1-2}
  \vspace{-5pt}
\end{figure} 
\begin{figure}
\centering
  \begin{subfigure}{0.9\columnwidth}
  \centering
    \begin{tikzpicture}
    \node(a){\includegraphics[width=0.5\linewidth]{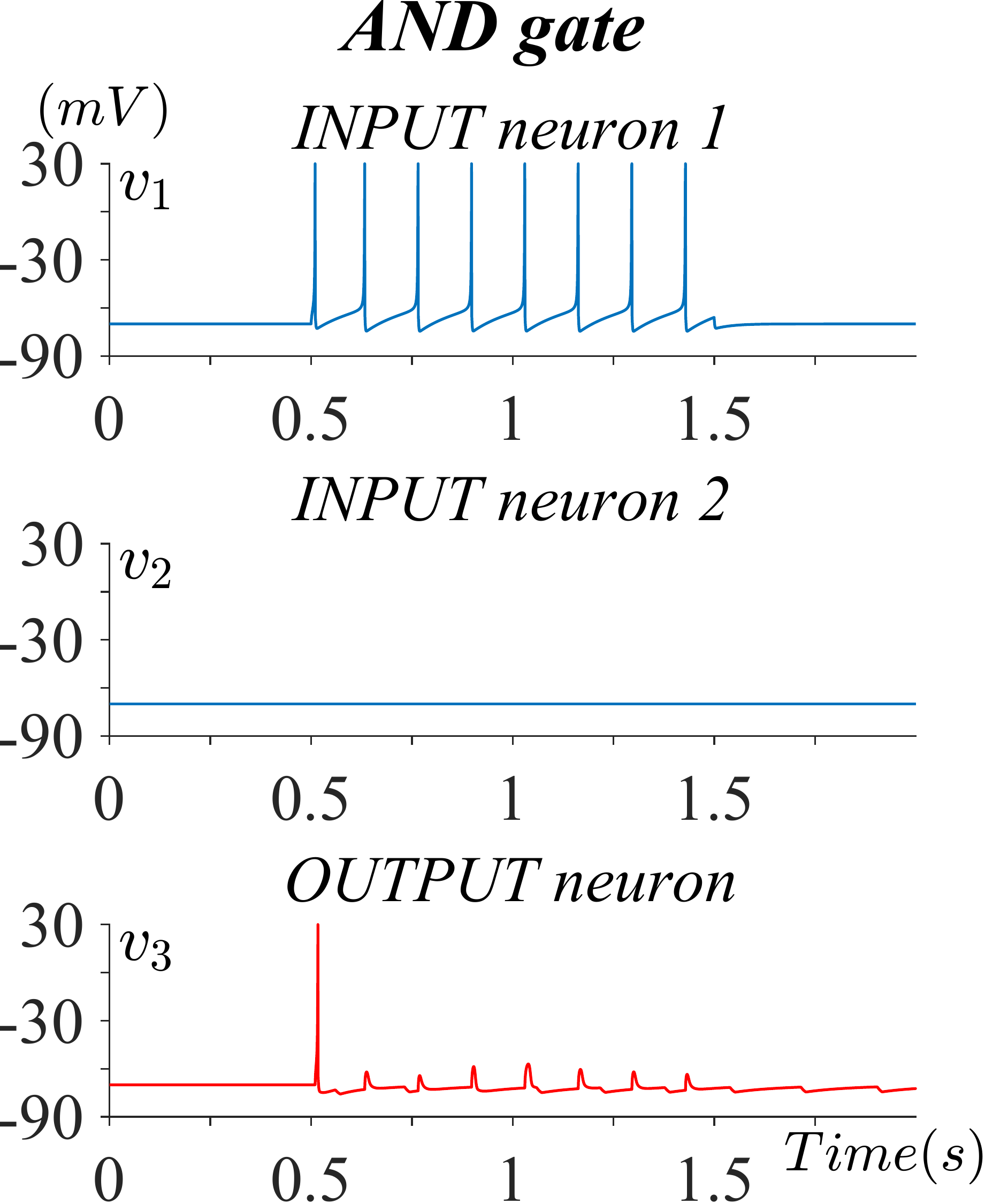}};
    \node at (a.north east)
    [
    anchor=center,
    xshift=+8mm,
    yshift=-13.5mm
    ]
    {
        \includegraphics[width=0.45\linewidth]{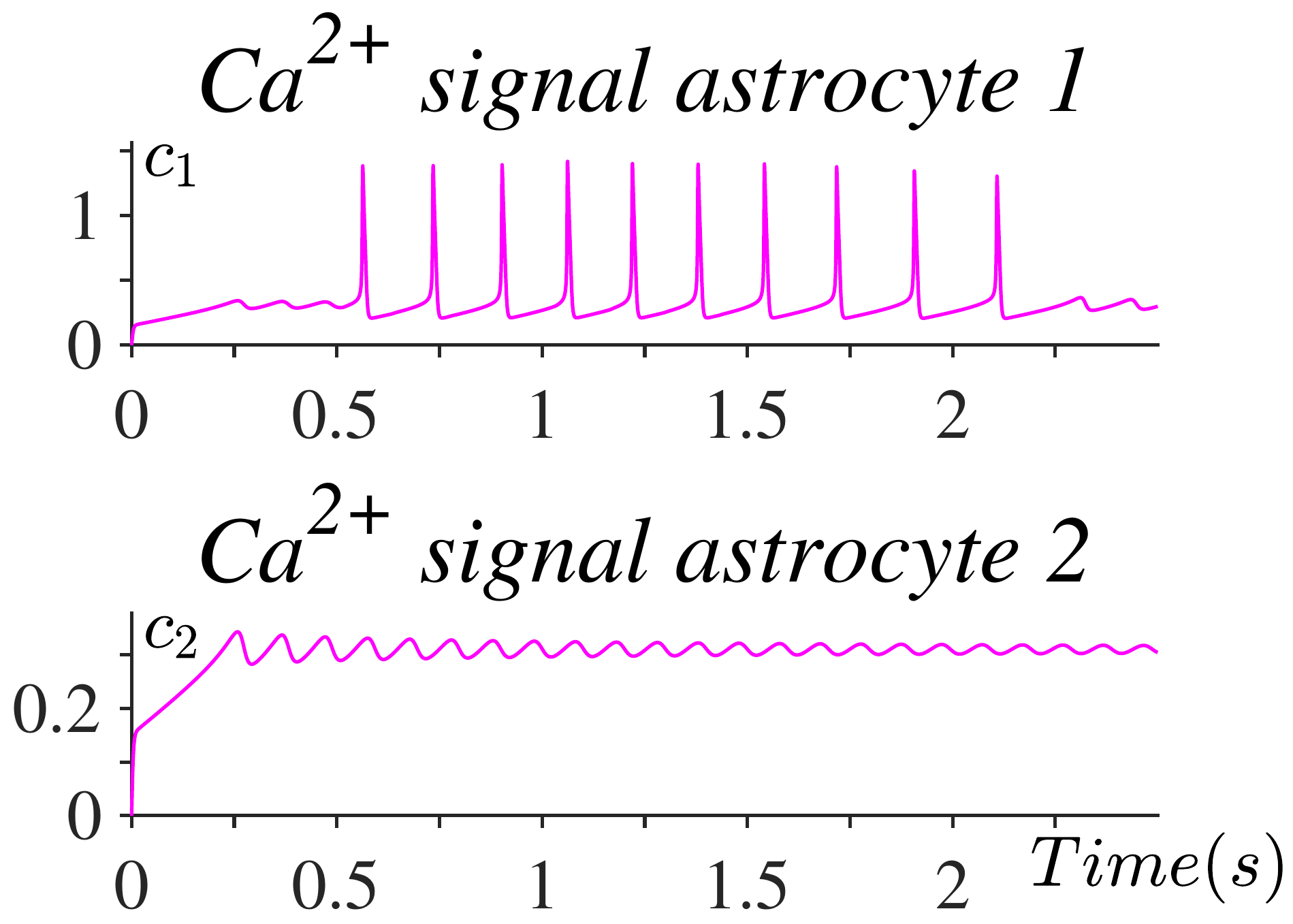}};
    \end{tikzpicture}
    \caption{inputs [1 0]}
    \label{figureB3a}
  \end{subfigure}
  \hfill 
  \\
  \begin{subfigure}{0.9\columnwidth}
  \centering
    \begin{tikzpicture}
    \node(a){\includegraphics[width=0.5\linewidth]{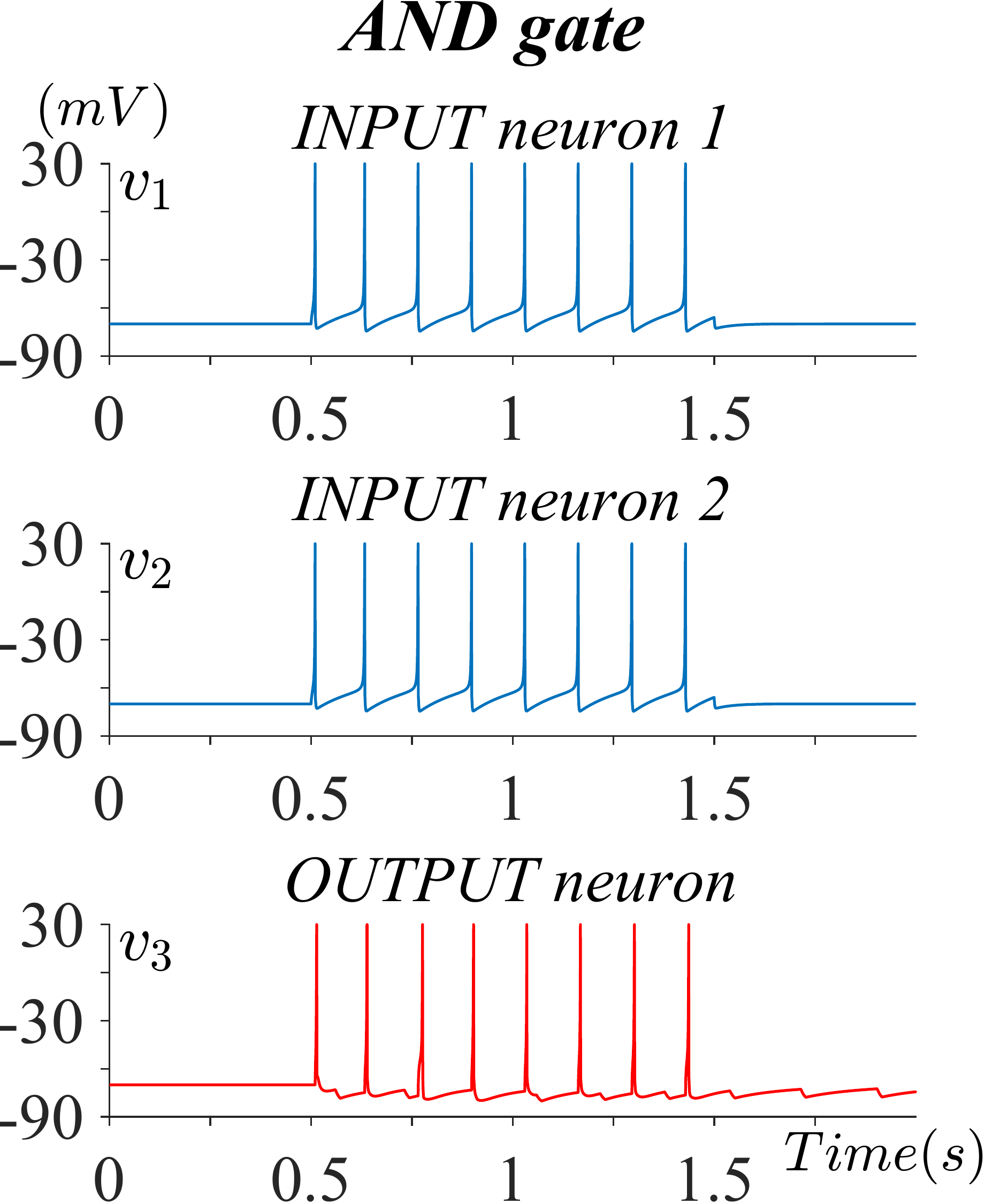}};
    \node at (a.north east)
    [
    anchor=center,
    xshift=+8mm,
    yshift=-13.5mm
    ]
    {
        \includegraphics[width=0.45\linewidth]{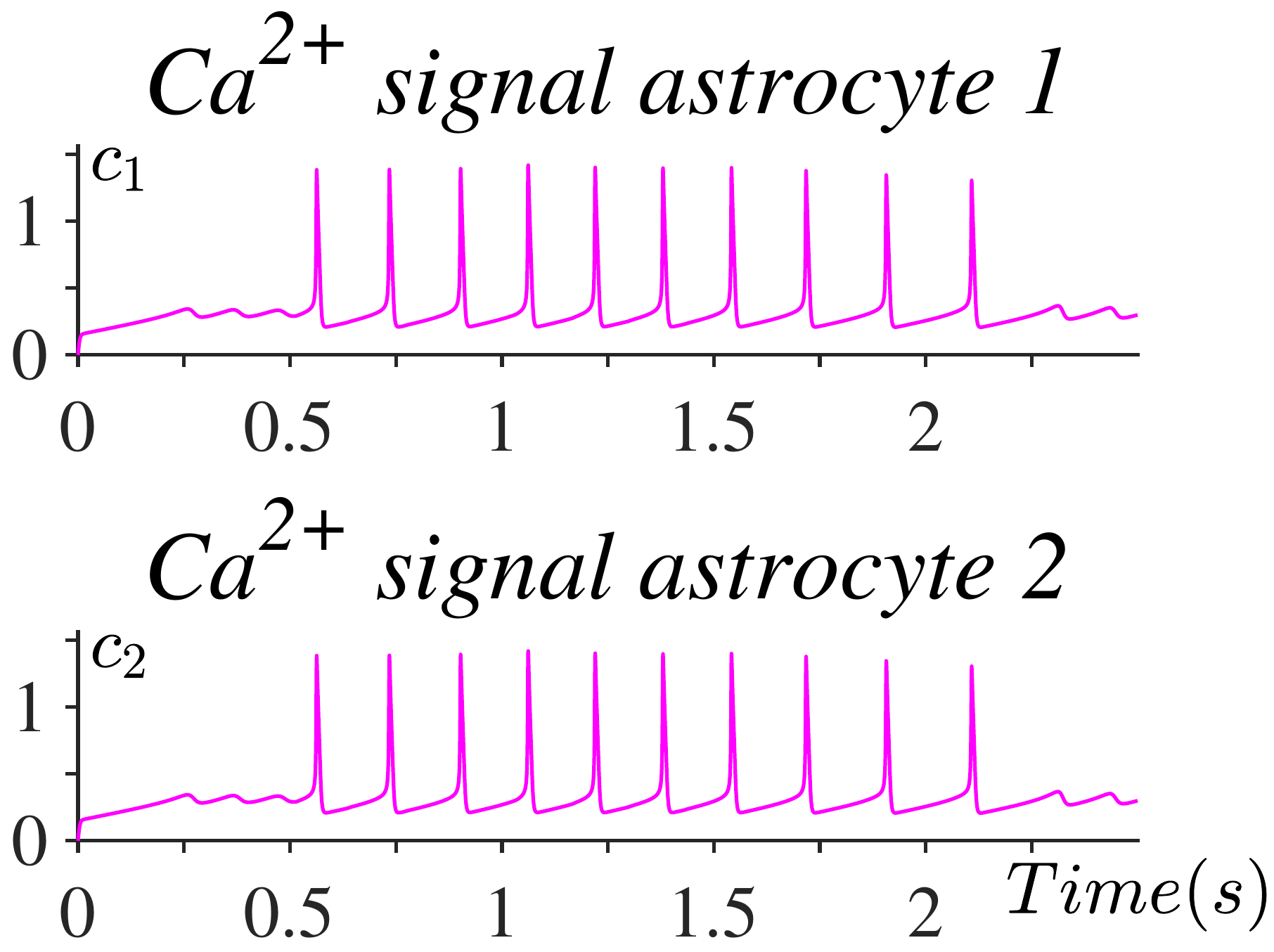}
    };
    \end{tikzpicture}
    \caption{inputs [1 1]}
  \end{subfigure}
  \caption{Simulation of AND gate with enabled astrocytes activity ($\alpha=0$ $\beta=0.05$ $\gamma=1.5$ $\delta=10$), synaptic strength $w_i=0.11$ and stimulating current $I=4$ pA. The astrocytes calcium signals are reported.}
  \label{B3-4}
  \vspace{-5pt}
\end{figure} 
In \cite{Postnov2007FunctionalMO} they showed how different calcium dynamics can be simulated by choosing astrocytes control parameters $\alpha$, $\beta$, $\gamma$, and $\delta$. In Fig. \ref{B3-4} an example of logic gates exploiting astrocytes activity is reported. Here, the fast activation pathway is blocked by setting $\alpha=0$, while the slow activation pathway is enabled with $\beta=0.05$. With this choice, the activity of each astrocyte reflects the state of the associated input neuron. When the input is not firing, astrocyte dynamic occurs as damped calcium oscillations, while when the input is firing, also the calcium signal shows spike activities. Moreover, astrocytes' activation continues also when neurons firing has already stopped. The astrocyte control parameters $\gamma$ and $\delta$, which reflect the positive feedback and negative feedback mechanism respectively, are empirically chosen so that the output neuron replicates the desired logic gating. Indeed these feedback effects can be used to modulate the influence of each stimulus in order to design different logic functions, similarly as previously done by changing the synaptic strengths.

\subsection{Effect of Noise in the Logic Gates}
\begin{figure*}[t]
\begin{minipage}[t]{.49\textwidth}
\centering
\subfloat[inputs {[1 0]}\label{C1}]{\includegraphics[width=.47\linewidth]{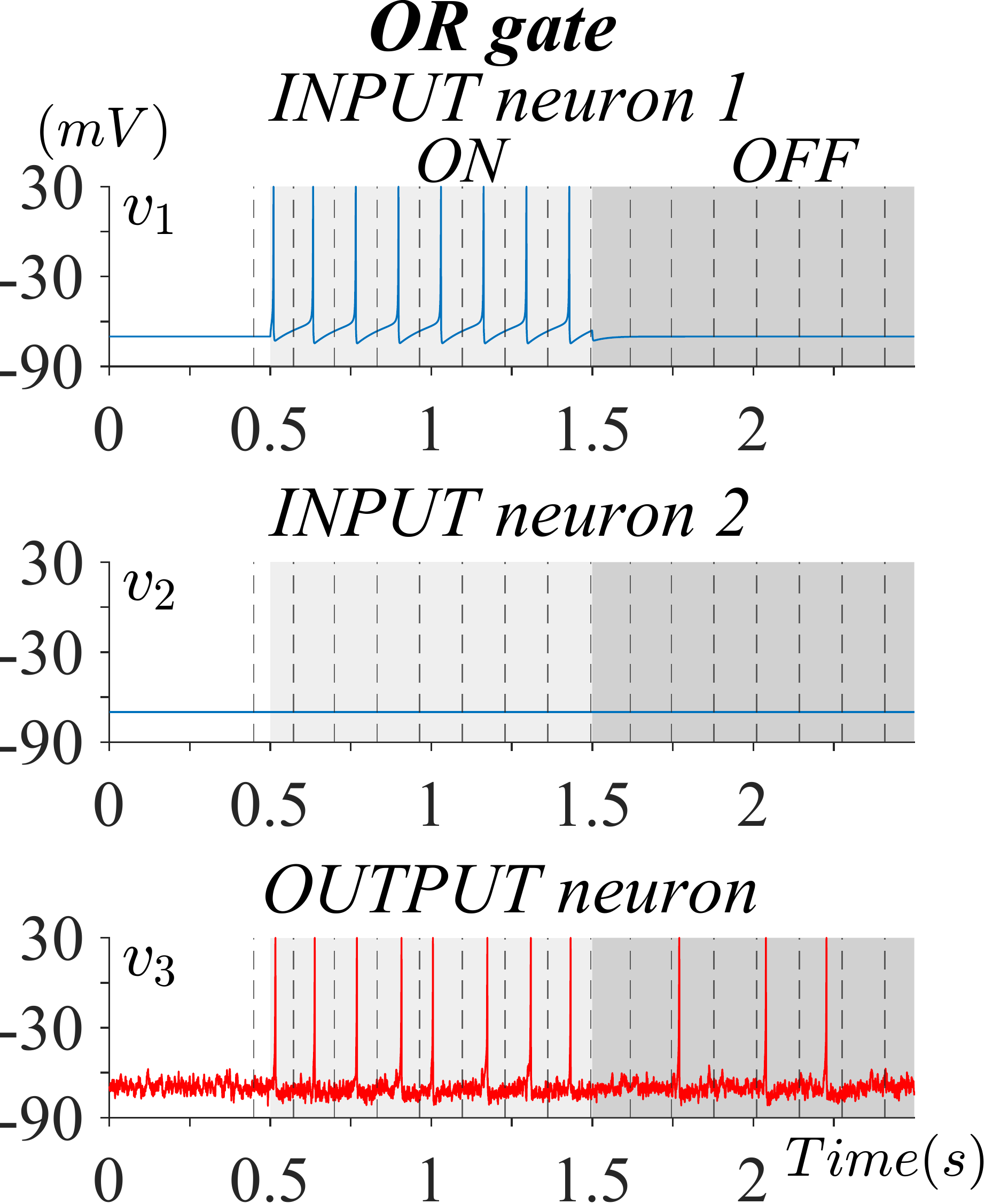}}\quad
\subfloat[inputs {[1 1]}\label{C2}]{\includegraphics[width=.47\linewidth]{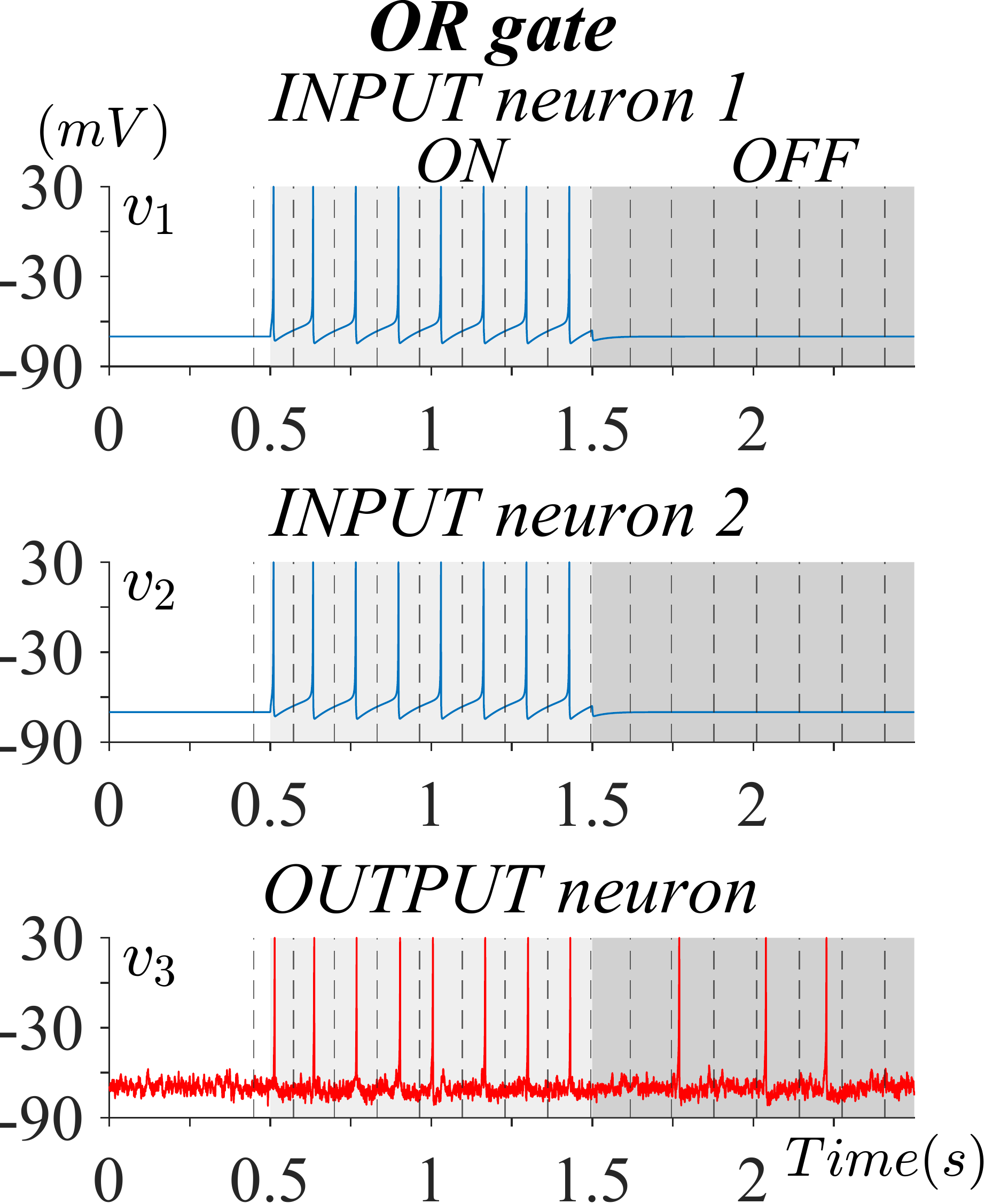}}
\caption{Simulation of OR gate involving only neurons, with synaptic strength $w_i=0.09$, stimulating current $I=4$ pA and affected by gaussian synaptic noise with $\sigma=5$. The stimulation current ON phase is marked with a light gray background, while the OFF phase with a dark gray background. The quality indexes values are: $\text{accuracy}=0.81$, $\text{LER}=18.75\%$ for both simulation \ref{C1} and \ref{C2}.}
\label{C1-2}
\end{minipage}\hfill
\vspace{2mm} 
\begin{minipage}[t]{.49\textwidth}
\centering
\subfloat[inputs {[1 0]}\label{C3}]{\includegraphics[width=.47\linewidth]{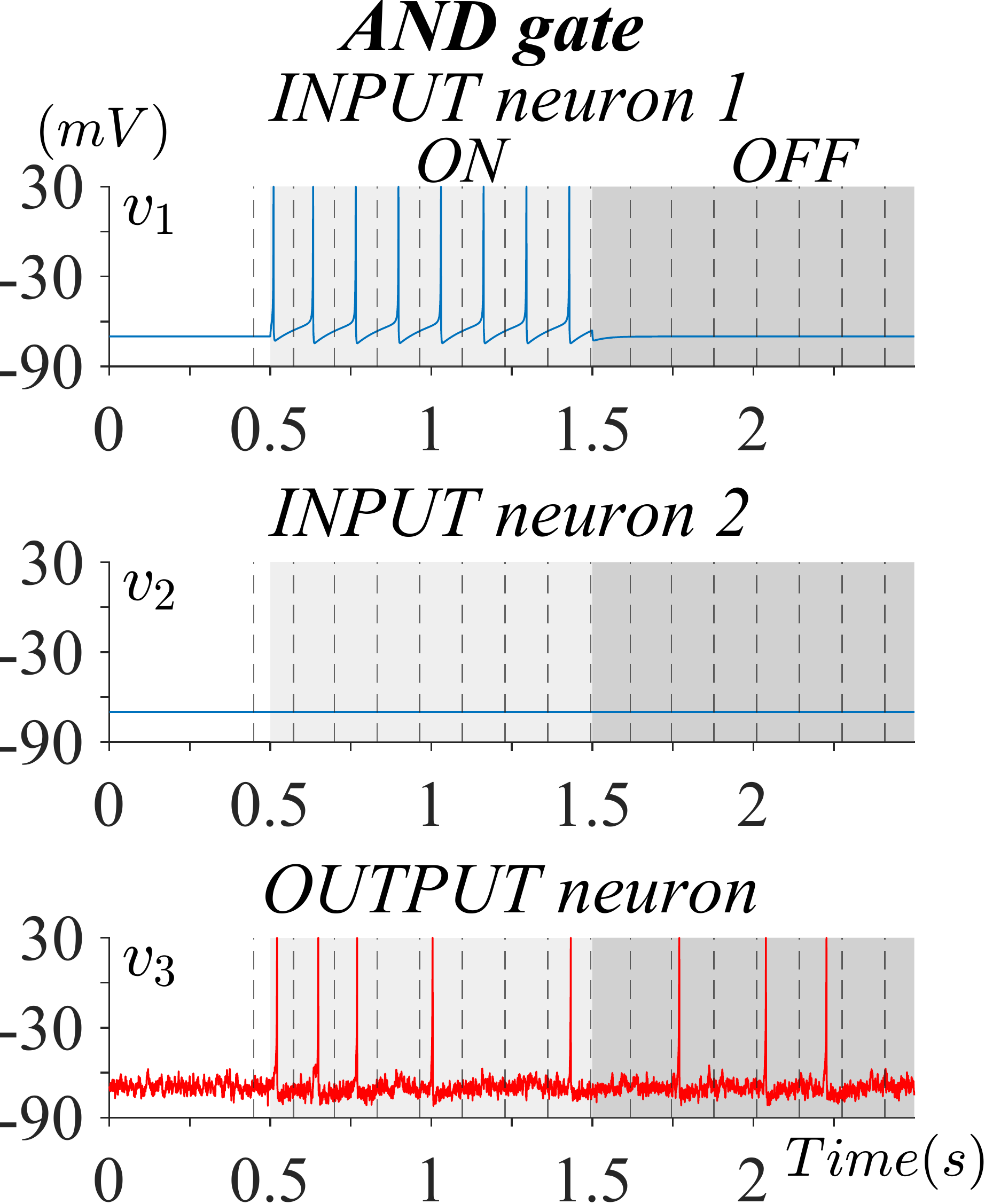}}\quad
\subfloat[inputs {[1 1]}\label{C4}]{\includegraphics[width=.47\linewidth]{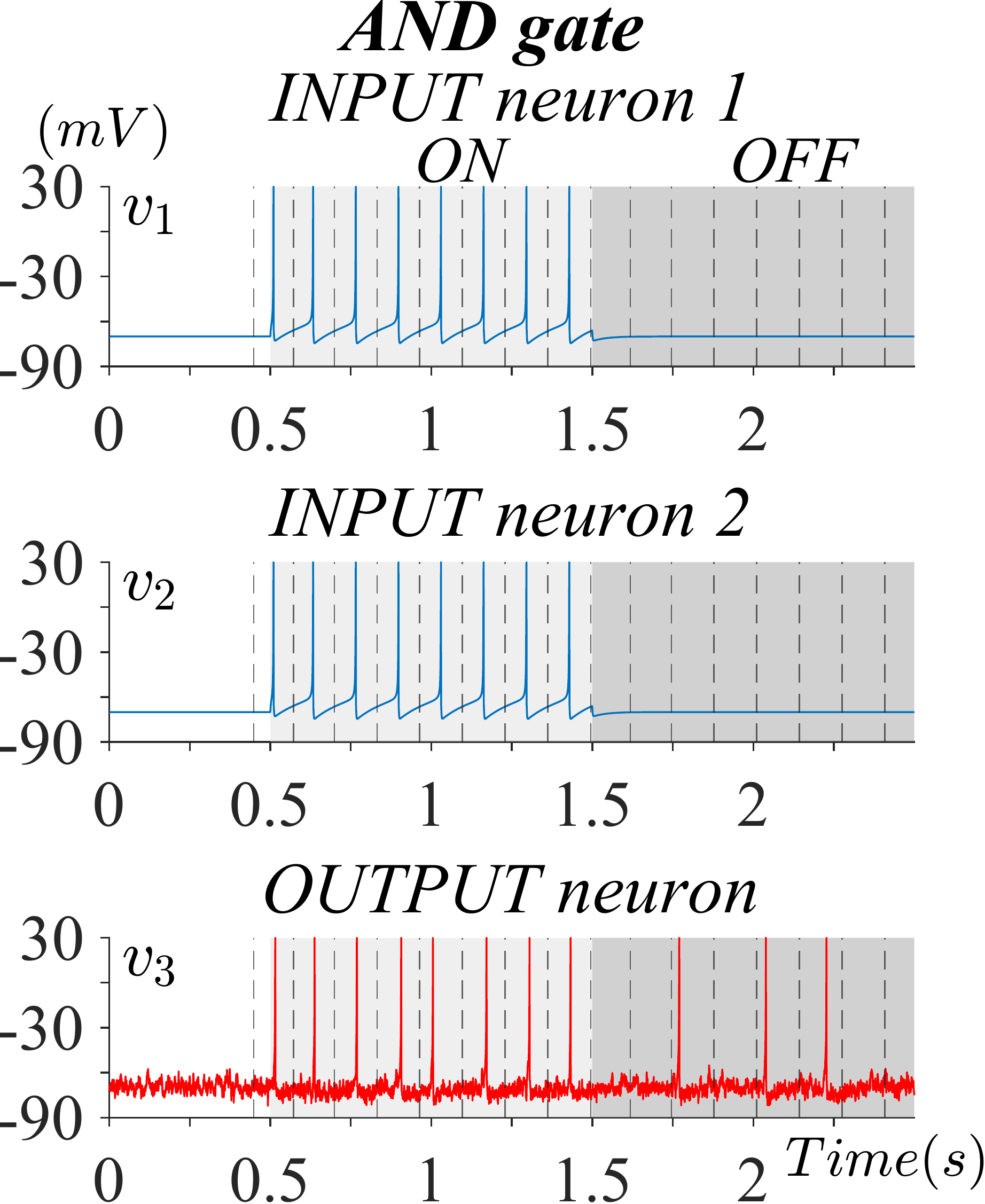}}
\caption{Simulation of AND gate involving only neurons, with synaptic strength $w_i=0.05$, stimulating current $I=4$ pA and affected by gaussian synaptic noise with $\sigma=5$. The stimulation current ON phase is marked with a light gray background, while the OFF phase with a dark gray background. The quality indexes values are: $\text{accuracy}=0.50$, $\text{LER}=50.00\%$ for simulation \ref{C3}, and $\text{accuracy}=0.81$, $\text{LER}=18.75\%$ for simulation \ref{C4}.}
\label{C3-4}
\end{minipage}\hfill
\begin{minipage}[t]{.49\textwidth}
\centering
\subfloat[inputs {[1 0]}\label{D1}]{\includegraphics[width=.47\linewidth]{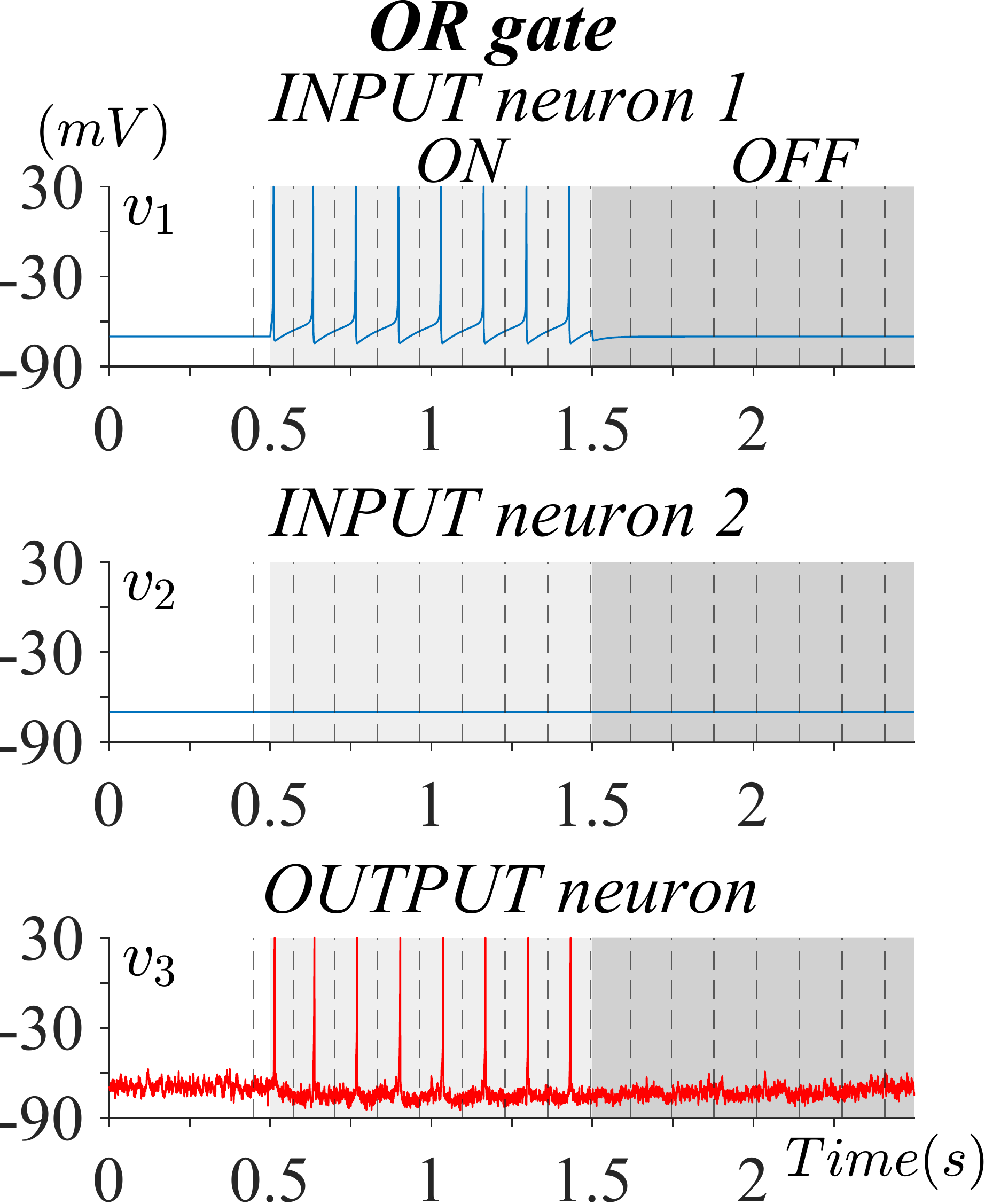}}\quad
\subfloat[inputs {[1 1]}\label{D2}]{\includegraphics[width=.47\linewidth]{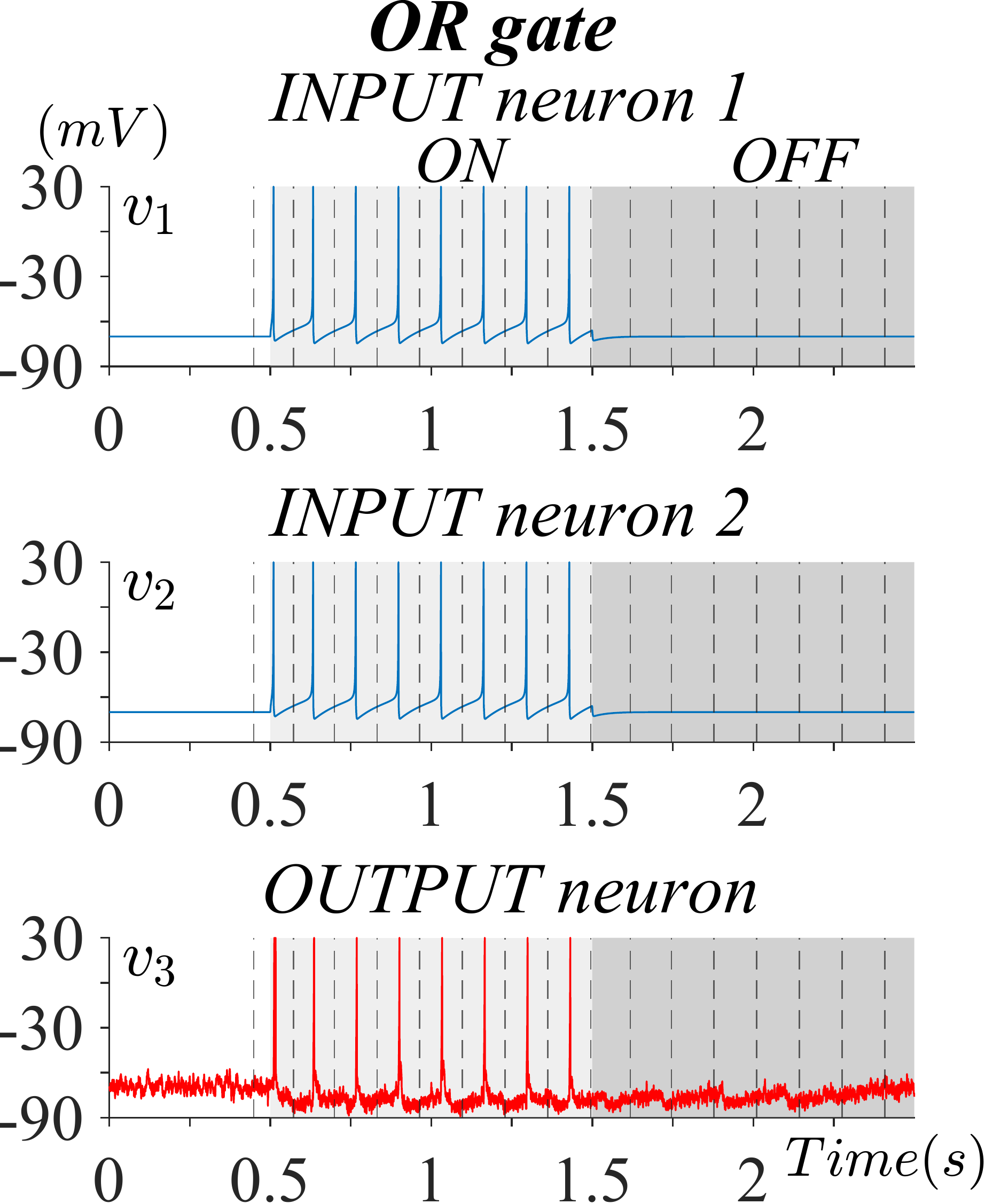}}
\caption{OR gate with astrocyte-based denoising. The synaptic strength is set as $w_i=0.22$, the stimulating current as $I=4$ pA, while the astrocyte control parameters are chosen as $\alpha=0$ $\beta=0.05$ $\gamma=0$ $\delta=15$. The stimulation current ON phase is marked with a light gray background, while the OFF phase with a dark gray background. The quality indexes values are: $\text{accuracy}=1.00$, $\text{LER}=0.00\%$ for simulation \ref{D1}, and $\text{accuracy}=0.94$, $\text{LER}=0.00\%$ for simulation \ref{D2}.}
\label{D1-2}
\end{minipage}\hfill
\begin{minipage}[t]{.49\textwidth}
\centering
\subfloat[inputs {[1 0]}\label{D3}]{\includegraphics[width=.47\linewidth]{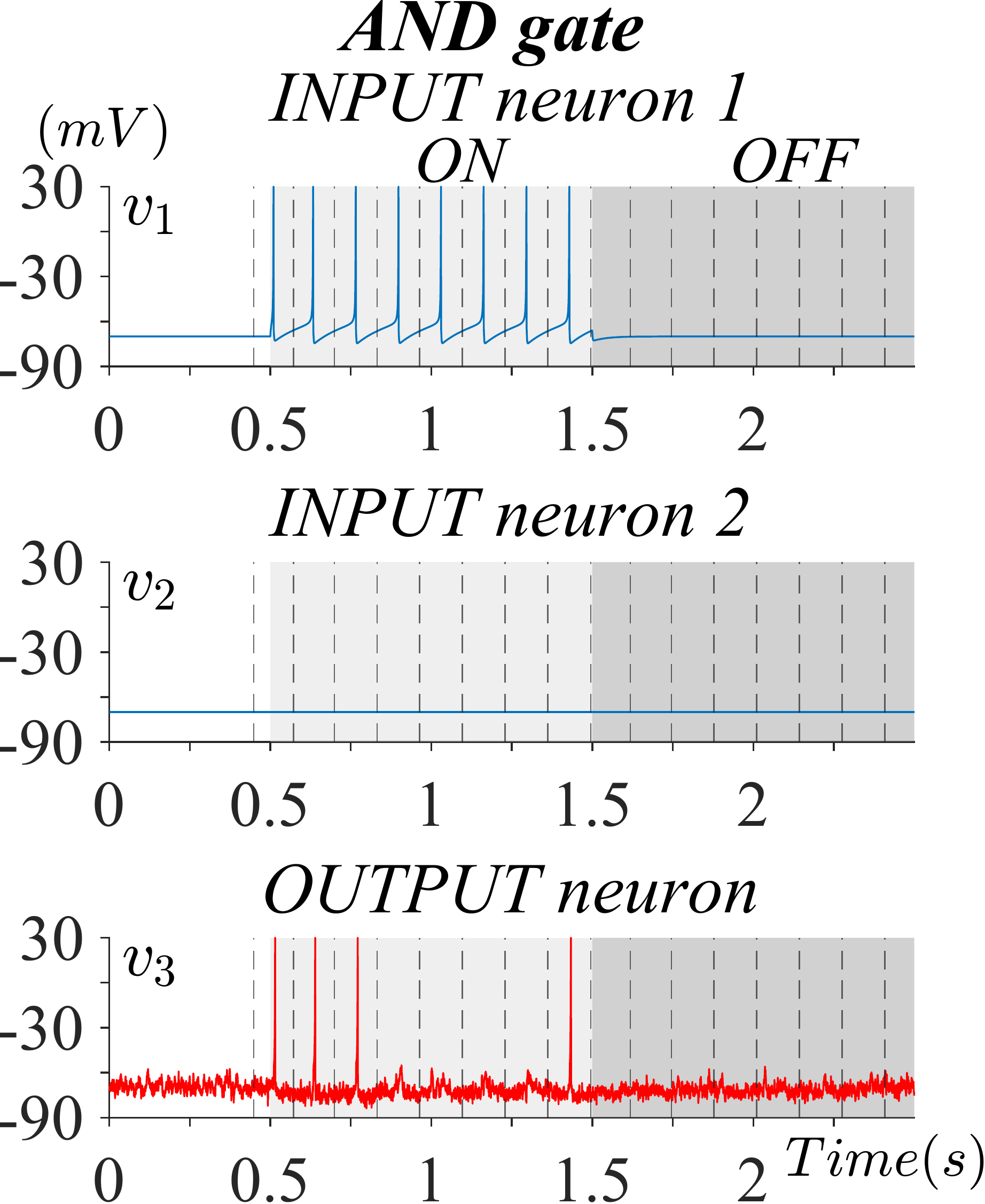}}\quad
\subfloat[inputs {[1 1]}\label{D4}]{\includegraphics[width=.47\linewidth]{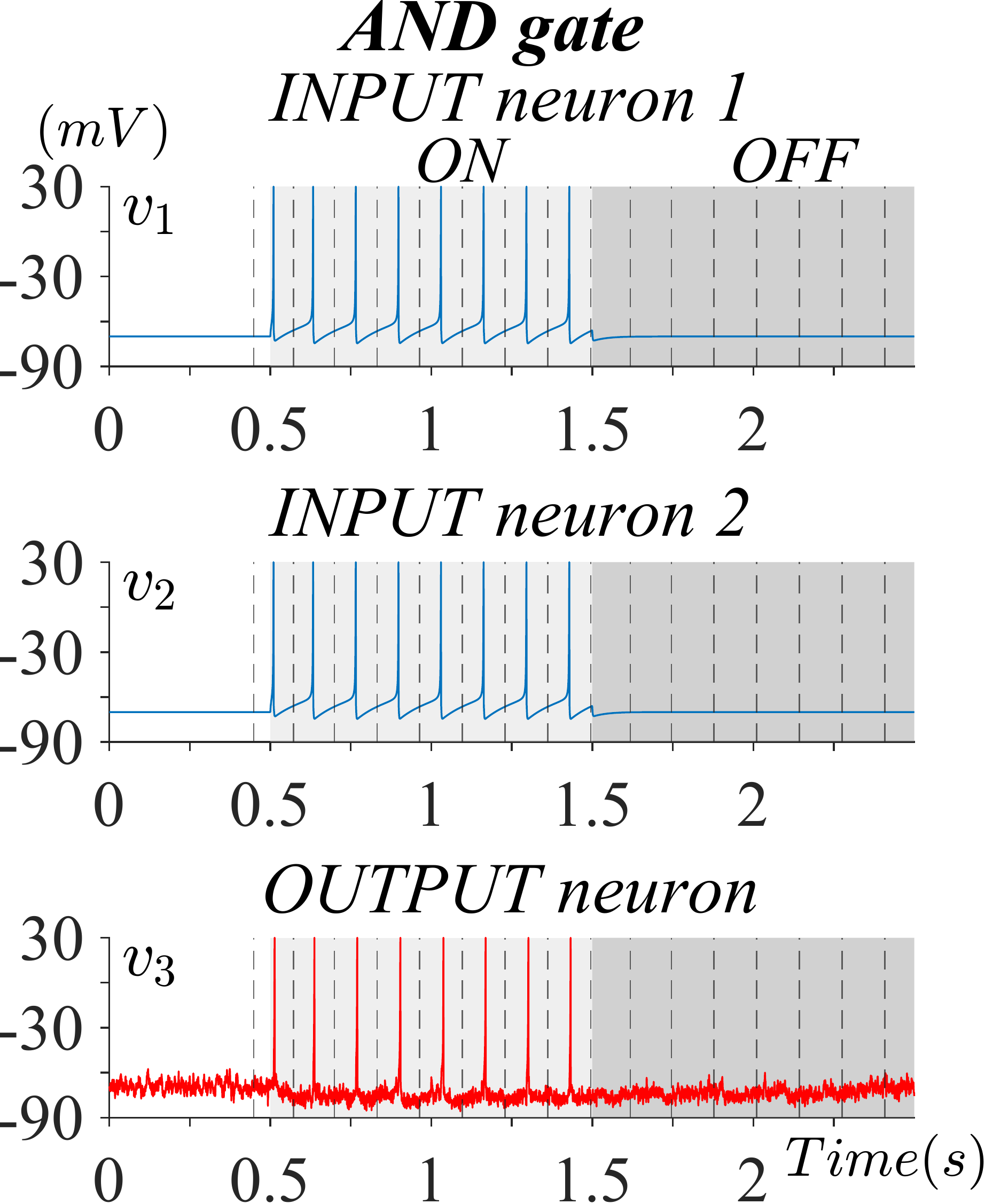}}
\caption{AND gate with astrocyte-based denoising. The synaptic strength is set as $w_i=0.11$, the stimulating current as $I=4$ pA, while the astrocyte control parameters are chosen as $\alpha=0$ $\beta=0.05$ $\gamma=1.5$ $\delta=10$. The stimulation current ON phase is marked with a light gray background, while the OFF phase with a dark gray background. The quality indexes values are: $\text{accuracy}=0.75$, $\text{LER}=25.00\%$ for simulation \ref{D3}, and $\text{accuracy}=1.00$, $\text{LER}=0.00\%$ for simulation \ref{D4}.}
\label{D3-4}
\end{minipage}
\vspace{-5pt}
\end{figure*}

\begin{figure*}[t]
\centering
  \begin{subfigure}[b]{0.75\columnwidth}
    \includegraphics[width=\linewidth]{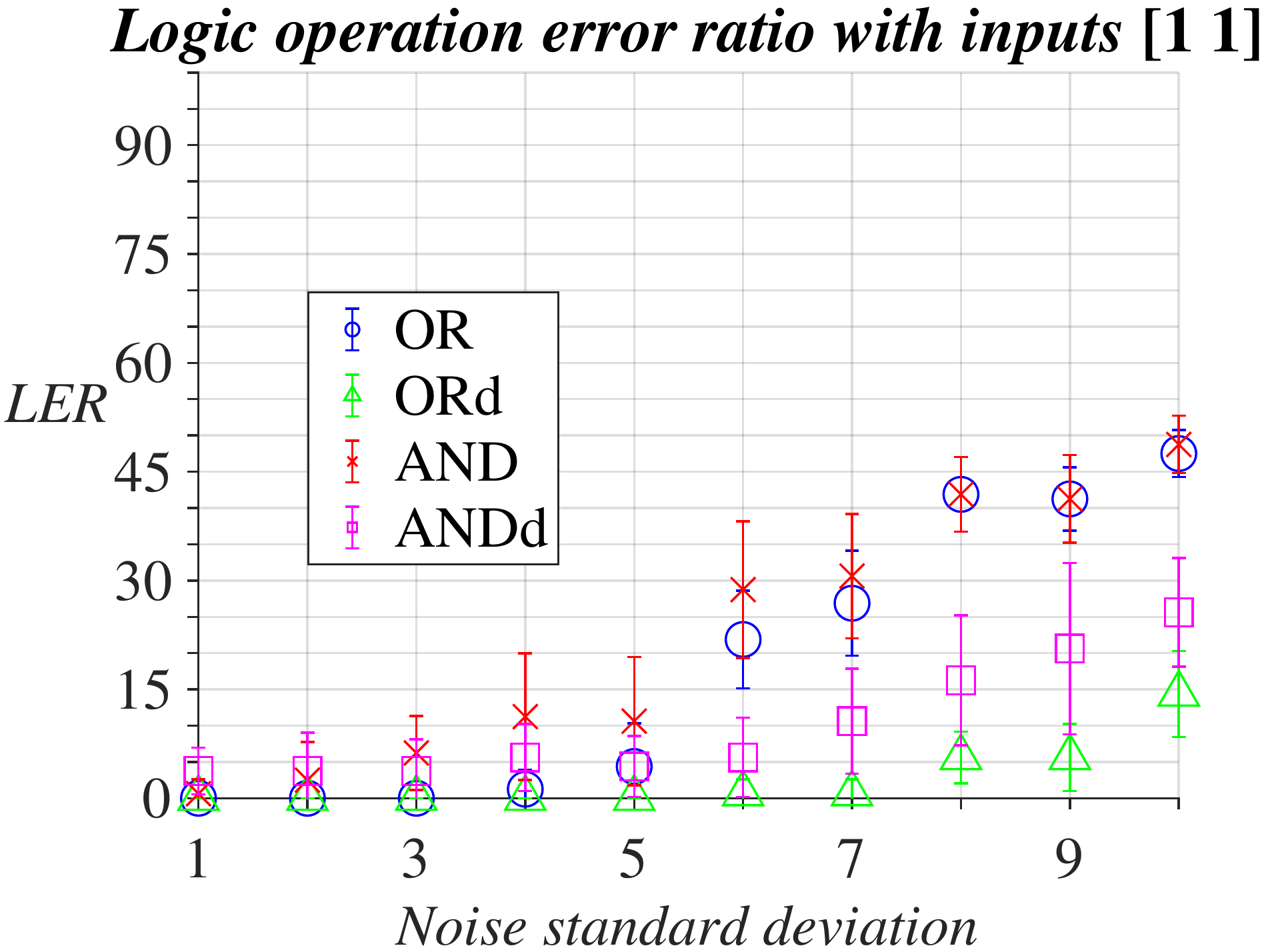}
    \caption{}
    \label{fig:1}
  \end{subfigure}
  \begin{subfigure}[b]{0.75\columnwidth}
    \includegraphics[width=\linewidth]{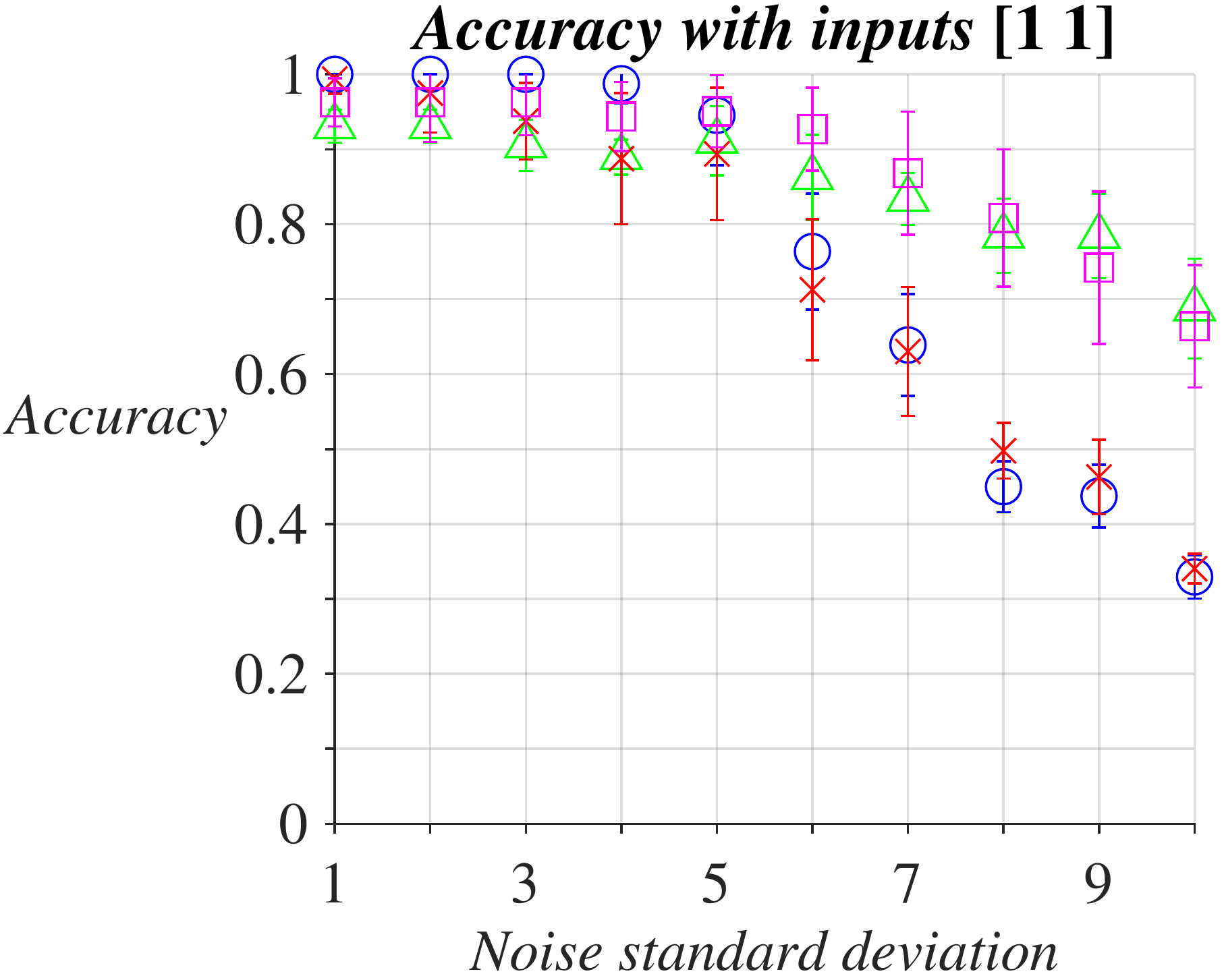}
    \caption{}
    \label{fig:2}
   \end{subfigure}
   \begin{subfigure}[b]{0.75\columnwidth}
    \includegraphics[width=\linewidth]{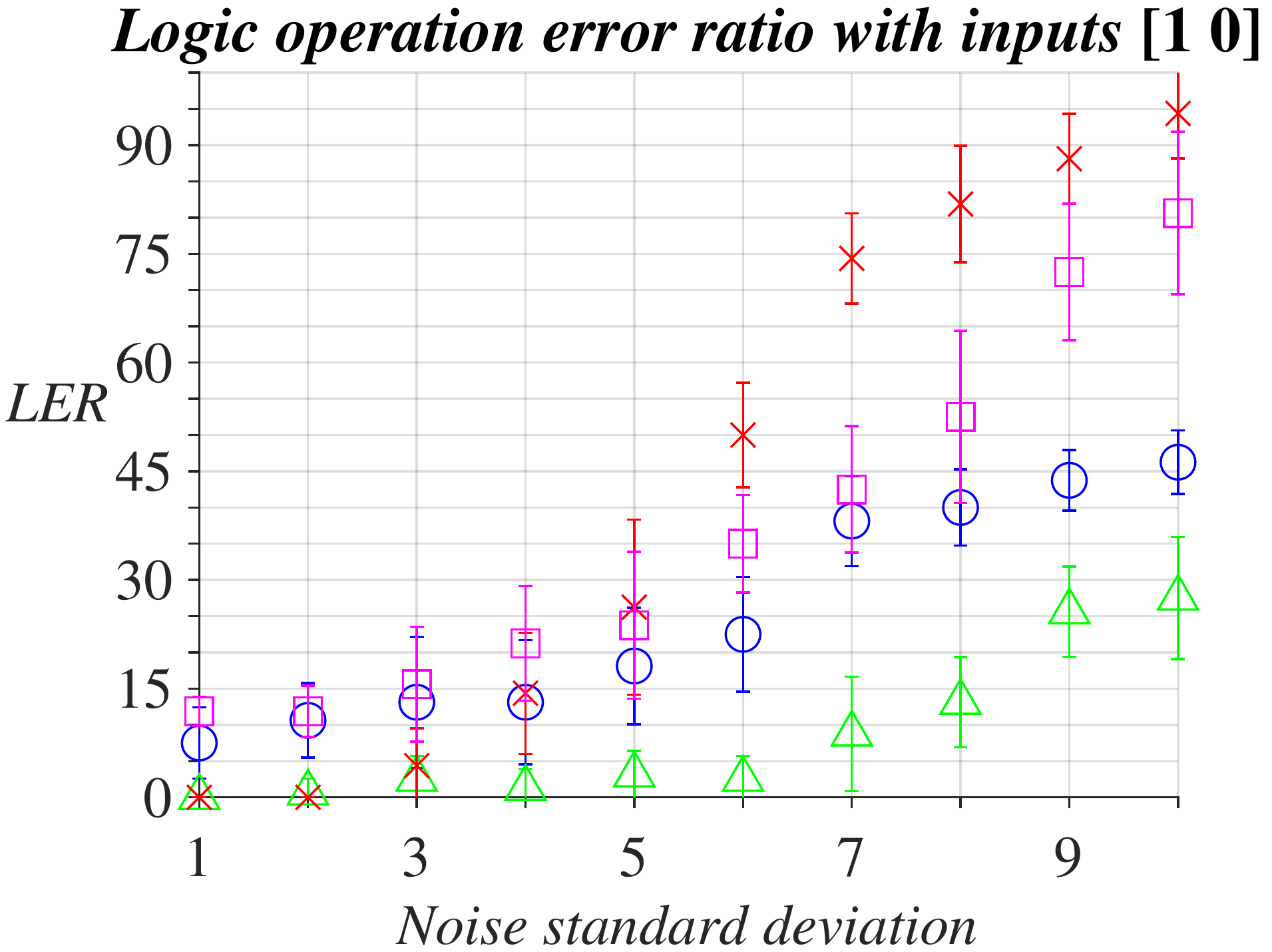}
    \caption{}
    \label{fig:3}
  \end{subfigure}
   \begin{subfigure}[b]{0.75\columnwidth}
    \includegraphics[width=\linewidth]{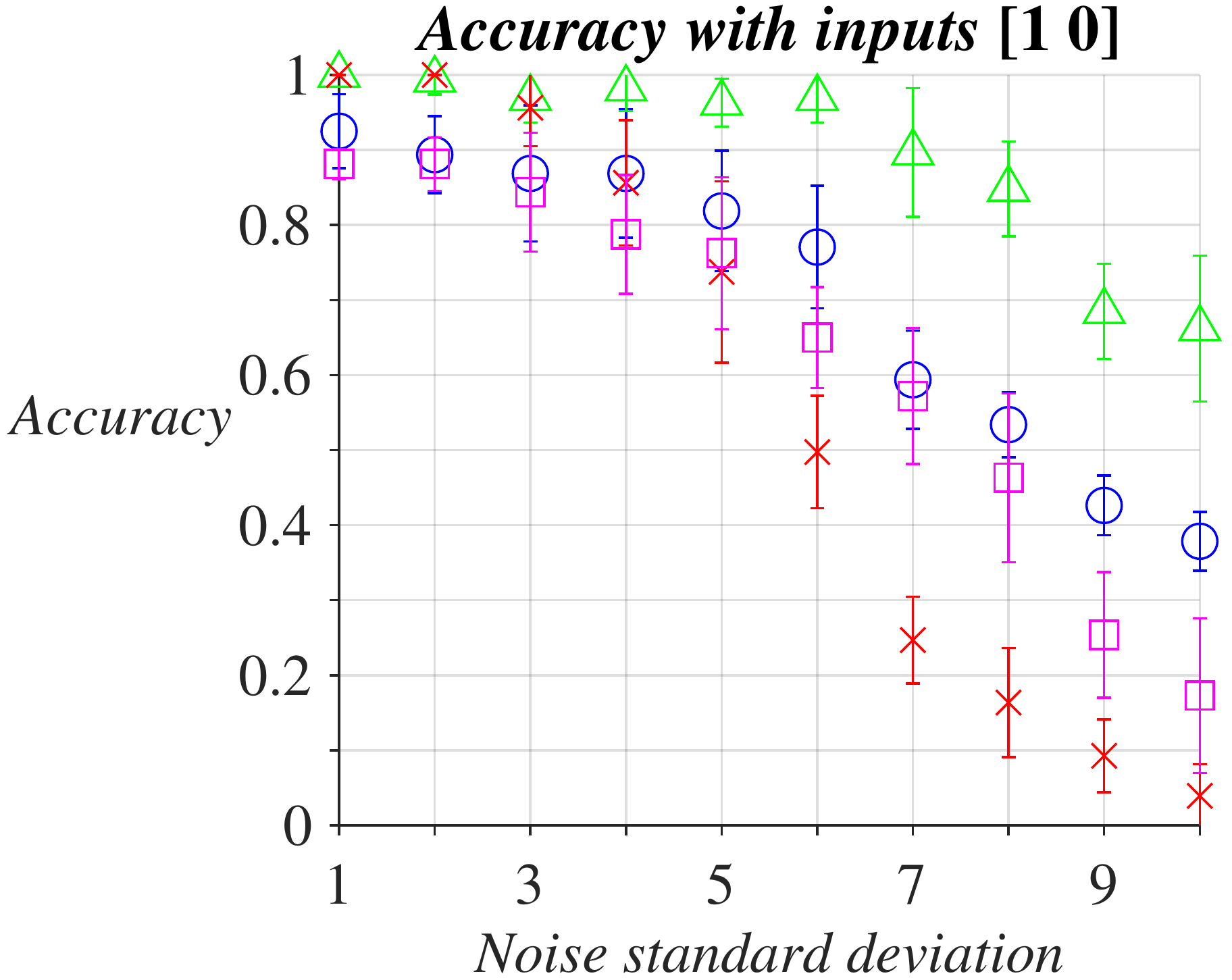}
    \caption{}
    \label{fig:4}
  \end{subfigure}
  \caption{Noise sensitivity analysis. Noise standard deviation changes between 1 to 10 and for each noise level 10 observations are generated. The graphs report the LER and accuracy averages and standard deviations of the output response for OR gate, OR gate with denoising mechanism (ORd), AND gate, and AND gate with denoising mechanism (ANDd).}
  \label{analysis}
  \vspace{-5pt}
\end{figure*}

The neuronal logic gates involving only neurons are tested with the presence of noise, using the noise model described in Section \ref{noise model}. The stimulating current is chosen as a rectangular function, with amplitude $I=4$ pA between 0.5 and 1.5 seconds (ON phase) and null between 1.5 and 2.5 seconds (OFF phase). All the noisy simulations reported in the following Figs. (from Fig. \ref{C1-2} to Fig. \ref{D3-4}), are generated using the same observation of synaptic noise, with standard deviation $\sigma=5$. The current OFF phases are highlighted using dark gray background, while the ON phases using light gray background. Neurons are modeled using the tonic spike pattern. 
To assess how noise influences the logic gating, first of all, the definition of a grid in which evaluating the response is needed. This is defined using one of the two input signals, that specifically needs to be at the high level. In the ON phase, the signals are segmented, dividing each interspike interval into halves. Since the duration of the ON and OFF phases is the same, in the OFF phase the segmentation previously defined for the ON phase is replicated. The first bin represents a special case. In order to define all bins as symmetric, with a spike at the center, the first bin precedes the first peak with an interval equal to half of the distance between the first and second peaks. An example of this signal segmentation can be observed in Fig. \ref{C1-2}. Each of the two phases consists of 8 bins, so the overall signal can be interpreted as a binary signal made of 16 bits. Note that the first 0.5 seconds, where no neurons are stimulated, are not assessed.
Once defined the grid, the output response accuracy is evaluated and signal errors are assessed using the logic operation error ratio (LER). In order to evaluate signal errors, first of all, the signals are encoded into a bit stream, classifying a spiking bin as 1, and a resting bin as 0. The logic operation error ratio is calculated as the number of wrong bits divided by the total number of transferred bits, expressed as a percentage. For binary classification problems, the accuracy can be defined as:
\begin{equation}
    accuracy = \frac{TP+TN}{TP+TN+FP+FN} \,,
\end{equation}
\noindent where $TP$ represents the number of true positives, $FP$ is the number of false positives, and $FN$ is the number of false negatives. For our purpose, the spiking state is considered the positive class, while the resting state is the negative class.  For instance, in the case of the OR gate with one or both inputs at the high level, since the output response needs to be at level 1, an output bin with a spike is considered a $TP$ while a bin without any spike is considered an $FN$. Given that spikes are assessed individually, if instead of one spike two spikes are found in the same bin, one spike counts as a $TP$ and the other as an $FP$. When both inputs are at level 0, the output response has to be 0, therefore an output bin without any spike is considered a $TN$, whereas a bin with one spike is considered an $FP$. 

Figs. \ref{C1-2} and \ref{C3-4} show an example of OR gate and AND gate with only neurons and affected by noise, with standard deviation $\sigma=5$.

\subsection{Astrocyte-based denoising}

In this Subsection, the possibility of reducing the effect of the synaptic noise using astrocytes regulation mechanisms is investigated. The underlying assumption is that the noise contribution is weaker than the stimulating inputs, and so the negative feedback mechanism should mainly affect the noise while its influence on the inputs should be neglectable. 
Fig. \ref{D1-2} presents the logic gate response of the OR gate with astrocytes-based denoising. The astrocytes control parameters are set as $\alpha=0$, $\beta=0.05$, $\gamma=0$ and $\delta=15$. Afterwards, the denoising mechanism is tested on the AND gate, as can be observed in Fig. \ref{D3-4}. Here the control parameters are chosen as $\alpha=0$, $\beta=0.05$, $\gamma=1.5$ and $\delta=10$. Since now both $\gamma$ and $\delta$ are nonzero, both positive and negative feedback are exploited.

Finally, the logic gates' performances are evaluated for increasing noise levels. The analysis is performed using noise standard deviation from 1 to 10, with 10 noise observations for each noise level. For each observation OR gate, OR gate with astrocyte denoising (ORd), AND gate, and AND gate with astrocyte denoising (ANDd) are assessed using accuracy and LER. The results of the noise sensitivity analysis are reported in Fig. \ref{analysis}.

\section{Discussion}
In Fig. \ref{B1-2} and \ref{B3-4} we can observe the effect of the astrocyte regulation mechanism and how it can be used to design different logic functions. These two figures are generated using the same network parameters, but with astrocytes' activity disabled (Fig. \ref{B1-2}) and enabled (Fig. \ref{B3-4}). In particular, an AND gate is designed using synaptic strength $w_i=0.11$, which is larger than the one used in Subsection \ref{Results A} ($w_i=0.05$). In Fig. \ref{B1-2}, with one input at level 1 and the other at 0, the output response is at level 1. Therefore the logic gate with only neurons does not correctly reproduce the AND function, due to the increased synaptic strength. Then the astrocytes negative feedback is used to correct the response, setting $\alpha=0$, $\beta =0.05$, $\gamma=1.5$ and $\delta =10$. As can be observed in Fig. \ref{B3-4}, the negative feedback mechanism is able to reduce the influence of the inputs and make the logic gate behaviour more similar to the AND logic function. Although
the overall response approximately follows the behaviour of the logic AND, in the case with inputs [1 0] (Fig. \ref{figureB3a}) the output shows an initial spike that should not be present. 
This can be caused by the fact that the astrocyte dynamic is slower than the neuron dynamic, and so the astrocyte regulation mechanism activates with a delay after the neuron excitation. Moreover, we exploited only the slow activation pathway, because the astrocyte feedback induced through the fast activation pathway was too strong. Despite that, the use of tripartite synapses results in significant benefits.
Indeed, we have tested the astrocyte regulating activity as a mechanism of reduction of synaptical noise. Figures \ref{C1-2} and \ref{C3-4} report the noisy simulations of logic gates employing only neurons. Most of these simulations have the same values of accuracy and LER, probably because the same noise observation was used. However, the AND gate with inputs [1 0] (Fig. \ref{C3}) is significantly more sensitive to noise with respect to the other cases, displaying the fall of the accuracy to 0.50 and the rise of the LER to 50.00\%. Generally, it has been observed that with the chosen synaptic noise model, noisy contributions often occur as spurious spikes rather than as the suppression of the true spikes. For this reason, in the case in which the logic response is high since the output neuron is already spiking, the noise effect could reinforce the spiking activity, but the errors introduced are not relevant to the interpretation of the binary message. By contrast, when the output needs to be at the resting state, as for the case of AND gate with inputs [1 0], the presence of noisy spikes significantly compromises the binary message. 

The astrocyte-based denoising mechanism is tested in Figs. \ref{D1-2} and \ref{D3-4}. When applied to the OR gate, the denoising method is able to noticeably reduce the noise contribution. Indeed, with inputs [1 0], the output perfectly matches the desired response (accuracy=1.00, LER=0.00\%). Instead, with inputs [1 1], the denoised OR gate displays slightly lower values of accuracy, because the first firing bin consists of two rapid spikes. However, this fact does not represent an effective binary error, since it reinforces the true response. The LER, which is not affected by additive true spikes, is still optimal (LER=0.00\%). As far as concerns the AND gate, even though the results with inputs [1 1] perfectly match the desired response, the case with inputs [1 0] remains a critical point. One possible interpretation could be related to the reduced synaptic strength. Indeed, since the AND gate $w_i$ is smaller than the OR one, the inputs that stimulate the postsynaptic neuron are less strong, and so it is difficult to set astrocyte inhibitory feedback without removing also the correct gating response.

The effectiveness of astrocyte denoising is further proved by Fig. \ref{analysis}, where the logic gates undergo increasing levels of noise and multiple observations are averaged. In the case of both inputs at level 1, AND and OR logic operation error ratio increases with noise levels, but remains below 55. Similarly as said before, noisy spikes during the high phases of the output signal do not cause relevant errors in the binary sequence, and so noise mainly affects the low phase. For this reason, the LER values remain controlled. Instead, the accuracy values decrease fast with the noise levels, because this index is also influenced by additive spurious spikes during the low phases. For both quality metrics, the performances with astrocyte denoising (ORd and ANDd) clearly overcome the realisations with only neurons (AND and OR). As far as concerned the case with inputs [1 0], generally the two metrics reflect worse results than the case of [1 1], because when the output must be low, both of the signal phases are at the low level, hence evidently affected by the presence of noisy spikes. The AND gate with inputs [1 0] confirms to have worse performances with respect to the other logic gates. The denoising mechanism is able to enhance LER and accuracy levels, but the ANDd remains still more sensitive to noise than the simple OR gate without denoising. By contrast, the denoising technique clearly improves the OR gate outcomes, obtaining the best results in terms of LER and accuracy.

Focused examinations of neuronal biocomputing systems with mathematical modeling will provide useful insight in how networks of biological cells could be engineered to display well-defined responses comparable to digital computations. In our work, we only explored the noisy dynamic of neuronal signalling with the model of synaptic gaussian noise. This phenomenon could be further characterised by the use of more complex and biologically plausible models of synaptic noise, for example with Poisson-distributed noise. In addition to the noisy dynamic, the reliability of neuronal biocomputing systems needs to be characterised by many other complex neuronal dynamics. Among them, the nervous system displays the dependence of distinct regions accounted for different physiological functions. Hence, this mechanism, called functional brain connectivity, is crucial for the comprehension of neuronal information processing since the inputs of certain regions of the network could also interfere with the outputs of distant regions. Furthermore, other open questions result from the fact that the ultimate networks' topology of neuronal populations is often unknown and it strictly depends on the neuronal activity due to the processes of neuronal plasticity. One possibility is to use topologically constrained solutions that manipulate the size of a neuron population and the subsequent connections between neurons and astrocytes \cite{girardin2022topologically}. The presented research shows lights in the neuronal network requirements to be considered by biotechnologists to realise biocomputing solutions either in-vitro or in-vivo. 

\section{Conclusion}
In this paper, we show how an improved reliable biocomputing solution can be achieved when AND and OR logic gates are built with neurons-astrocytes networks. We present a spike pattern generic tripartite synapse model by coupling with modifications to the Izhikevich model and the Postnov model. Moreover, we expanded the model for the case of multiple tripartite synapses connected with the same postsynaptic neuron. We showed the realization of both AND and OR gates, characterized their performance in relation to noise, and quantified the denoising effects of astrocytes. An up to 25\% overall performance improvement in the logic gate reliability shows the positive astrocytes' contribution to computing tasks performed by neurons. We believe such modeling brings to light the role of other non-neuronal cells in the behaviour of logic gates, which is far from complete in this exciting area of biocomputing. We hope these solutions can be used for living implants or in-vitro solutions in the future.

\section*{Acknowledgement}

The work of M.T.B. is funded by the European Union's Horizon 2020 Research and Innovation Programme under the Marie Skłodowska-Curie grant agreement No. 839553.

\bibliographystyle{IEEEtran}
\bibliography{ref} 





\end{document}